\documentstyle[12pt,psfig]{article}
\newcommand{\nsect}{\setcounter{equation}{0}
\def\theequation{\thesection.\arabic{equation}}\section}

\newcommand{\appendixA}{\setcounter{equation}{0}
\def\theequation{\rm{A}.\arabic{equation}}\section*}
\newcommand{\appendixB}{\setcounter{equation}{0}
\def\theequation{\rm{B}.\arabic{equation}}\section*}

\catcode`\@=11
\def\marginnote#1{}
\newcount\hour
\newcount\minute
\newtoks\amorpm
\hour=\time\divide\hour by60
\minute=\time{\multiply\hour by60 \global\advance\minute by-
\hour}
\edef\standardtime{{\ifnum\hour<12 \global\amorpm={am}%
    \else\global\amorpm={pm}\advance\hour by-12 \fi
    \ifnum\hour=0 \hour=12 \fi
    \number\hour:\ifnum\minute<100\fi\number\minute\the\amorpm}}
\edef\militarytime{\number\hour:\ifnum\minute<100\fi\number\minute}
\def\draftlabel#1{{\@bsphack\if@filesw {\let\thepage\relax
  \xdef\@gtempa{\write\@auxout{\string
    \newlabel{#1}{{\@currentlabel}{\thepage}}}}}\@gtempa
    \if@nobreak \ifvmode\nobreak\fi\fi\fi\@esphack}
     \gdef\@eqnlabel{#1}}
\def\@eqnlabel{}
\def\@vacuum{}
\def\draftmarginnote#1{\marginpar{\raggedright\scriptsize\tt#1}}
\def\draft{\oddsidemargin -.5truein
        \def\@oddfoot{\sl preliminary draft \hfil
        \rm\thepage\hfil\sl\today\quad\militarytime}
        \let\@evenfoot\@oddfoot \overfullrule 3pt
        \let\label=\draftlabel
        \let\marginnote=\draftmarginnote

\def\@eqnnum{(\theequation)\rlap{\kern\marginparsep\tt\@eqnlabel}%
\global\let\@eqnlabel\@vacuum}  }
\def\preprint{\twocolumn\sloppy\flushbottom\parindent 1em
        \leftmargini 2em\leftmarginv .5em\leftmarginvi .5em
        \oddsidemargin -.5in    \evensidemargin -.5in
        \columnsep 15mm \footheight 0pt
        \textwidth 250mmin      \topmargin  -.4in
        \headheight 12pt \topskip .4in
        \textheight 175mm
        \footskip 0pt

\def\@oddhead{\thepage\hfil\addtocounter{page}{1}\thepage}
        \let\@evenhead\@oddhead \def\@oddfoot{} \def\@evenfoot{}
}
\def\titlepage{\@restonecolfalse\if@twocolumn\@restonecoltrue\onecolumn
     \else \newpage \fi \thispagestyle{empty}\c@page\z@
        \def\thefootnote{\fnsymbol{footnote}} }
\def\endtitlepage{\if@restonecol\twocolumn \else  \fi
        \def\thefootnote{\arabic{footnote}}
        \setcounter{footnote}{0}}  
\catcode`@=12
\relax
\def\be{\begin{equation}}
\def\ee{\end{equation}}
\def\bea{\begin{eqnarray}}
\def\eea{\end{eqnarray}}
\def\simlt{\stackrel{<}{{}_\sim}}
\def\simgt{\stackrel{>}{{}_\sim}}

\def\NPB#1#2#3{{\it Nucl.~Phys.} {\bf{B#1}} (19#2) #3}
\def\PLB#1#2#3{{\it Phys.~Lett.} {\bf{B#1}} (19#2) #3}
\def\PRD#1#2#3{{\it Phys.~Rev.} {\bf{D#1}} (19#2) #3}
\def\PRL#1#2#3{{\it Phys.~Rev.~Lett.} {\bf{#1}} (19#2) #3}

\def\PTP#1#2#3{{\it Prog.~Theor.~Phys.} {\bf#1}  (19#2) #3}

\def\mst11{m_{\;\widetilde{t}_{1}}}
\def\msti{m_{\;\widetilde{t}_i}}
\def\mstj{m_{\;\widetilde{t}_j}}
\def\msbi{m_{\;\widetilde{b}_i}}
\def\msbj{m_{\;\widetilde{b}_j}}
\def\st{\;\widetilde{t}}
\def\sb{\;\widetilde{b}}
\def\mst22{m_{\;\widetilde{t}_{2}}}
\def\mst12{m_{\;\widetilde{t}_{1,2}}}
\def\mstl{m_{\;\widetilde{t}_L}}
\def\mstr{m_{\;\widetilde{t}_R}}
\def\msb11{m_{\;\widetilde{b}_{1}}}
\def\msb22{m_{\;\widetilde{b}_{2}}}
\def\msb12{m_{\;\widetilde{b}_{1,2}}}
\def\msbl{m_{\;\widetilde{b}_L}}
\def\msbr{m_{\;\widetilde{b}_R}}
\def\modh{\left|H\right|}
\def\mtilde2{\widetilde{m}^{2}}
\def\lambdatilde{\widetilde{\lambda}}
\def\Lambdatilde{\widetilde{\Lambda}}
\def\leff{\lambda_{\rm eff}}
\def\mbart{\overline{m}_{t}}
\def\exis{\varphi}
\relax

\begin{document}
\topmargin-2.5cm
%
\begin{titlepage}
\begin{flushright}
CERN-TH/95-157\\
IEM-FT-106/95 \\
hep--ph/9508343 \\
\end{flushright}
\vskip 0.3in
\begin{center}{\Large\bf
Effective potential methods and  \\
the Higgs mass spectrum in the MSSM
\footnote{Work supported in part by
the European Union (contract CHRX-CT92-0004) and
CICYT of Spain
(contract AEN94-0928).}  } \\
\vskip .5in
{\bf M. Carena},
{\bf M. Quir\'os}
\footnote{On leave of absence from Instituto
de Estructura de la Materia, CSIC, Serrano 123, 28006-Madrid,
Spain.}
and {\bf C.E.M. Wagner} \vskip.35in
CERN, TH Division, CH--1211 Geneva 23, Switzerland\\
\end{center}
\vskip2.cm
\begin{center}
{\bf Abstract}
\end{center}
\begin{quote}
We generalize the analytical expressions for the two-loop
leading-log neutral Higgs boson masses and mixing angles to the
case of  general left- and right-handed soft supersymmetry
breaking stop and sbottom masses and left--right mixing mass
parameters ($m_Q, m_U, m_D, A_t, A_b$). This generalization
is essential for the computation of Higgs masses and
couplings in the presence of light stops.
At high scales we use the minimal supersymmetric standard
model effective potential, while at low scales
we consider the two-Higgs doublet model (renormalization group
improved) effective potential, with general
matching conditions at the thresholds
where the squarks decouple.
We define physical (pole) masses for the top-quark, by including
QCD self-energies, and for the neutral Higgs bosons, by including
the leading one-loop electroweak self-energies where the top/stop and
bottom/sbottom sectors propagate.
For $m_Q = m_U = m_D$ and  moderate left--right mixing mass
parameters, for which the mass expansion in terms of
renormalizable Higgs quartic couplings is reliable,
we find excellent agreement with  previously obtained results.
\end{quote}
\vskip1.cm

\begin{flushleft}
CERN-TH/95-157\\
August 1995 \\
\end{flushleft}

\end{titlepage}
\setcounter{footnote}{0}
\setcounter{page}{0}
\newpage
%

\nsect{Introduction}

The use of effective potential methods has
proved to be an elegant and simple
way of incorporating all dominant top mass dependent one-loop
radiative corrections to the Higgs masses and mixing angles
in the minimal supersymmetric Standard Model
(MSSM)~\cite{topc,ERZ}. Indeed, the
one-loop effective potential computation reproduces
with a good level of accuracy the complete one-loop radiative
corrections to the Higgs masses~\cite{Marc} and
mixing angles.
However, relevant corrections are missing within this approximation.
In particular, for the values of the top-quark mass preferred by
the most recent experimental data~\cite{CDF,D0},
the next-to-leading order Higgs mass values
can differ from the leading order ones by 5 to 15 GeV.

The two-loop corrections to the Higgs masses in the limit of
large values of the CP-odd Higgs mass and  stop mass parameters
have been computed in different
approximations~\cite{EQ1}--\cite{CEQR}. In particular,
a complete computation of the
next-to-leading order effects
on the lightest Higgs mass was presented
in Ref.~\cite{CEQR}. It was
subsequently realized that, when the leading-log
renormalization group improved Higgs mass
expressions~\cite{LL}--\cite{HH}
are evaluated at the pole top-quark mass scale $M_t$, they
reproduce the next-to-leading order values
with a high level of accuracy, for
any value of $\tan\beta$ and of the stop mixing
mass parameter~\cite{CEQW}.
This means that, for this particular value of the renormalization
scale, the genuine two-loop corrections are small, an observation
analogous to the one already made in Ref.~\cite{EQ1}.
The same holds if the renormalization scale is fixed at the
on-shell top-quark mass, $\mbart=m_t(m_t)$, where $m_t$
is the running top-quark mass\footnote{The relation
between the running mass $m_t$ and pole
mass $M_t$, for the top-quark, taking into account
one-loop QCD corrections, is:
${\displaystyle M_t=m_t(M_t)
\left[1+\frac{4\alpha_3(M_t)}{3\pi}\right] }$.}.
Based on the above observation
about the choice of the renormalization scale,
we presented in Ref.~\cite{CEQW} analytical expressions
for the two-loop leading order Higgs masses and couplings, which
are valid for
common values of the stop and sbottom
supersymmetry breaking masses
and moderate values of the squark left--right mixing mass
parameters~\cite{CEQW}.
Indeed,  all next-to-leading
order computations of  the Higgs masses and couplings assumed
the left-handed and right-handed stop mass parameters to be
equal. Moreover, they relied on an expansion of the effective
potential which becomes reliable only for moderate values
of the stop mixing parameter (see the discussion in Ref.~\cite{CEQW}).
It is useful to find a good analytical approximation, which
works  independently of the nature of the stop spectrum. The main objective
of this work is to provide such an  analytical approximation, based
on an analysis of the dominant leading
order effects in the effective potential
computation of the neutral CP-even Higgs masses.

We now present, for completeness, the neutral Higgs one-loop
effective potential in the MSSM and define our
notation and conventions.  In a mass-independent renormalization scheme
such as $\overline{{\rm MS}}$ or $\overline{{\rm DR}}$~\cite{DR},
the one-loop effective potential of the MSSM, as a function of the neutral
components $H_1$ and $H_2$ of the two-Higgs doublets, is given by,
\be
\label{potgeneral}
V^{(1)} =\frac{3}{32\pi^2}\left\{
\sum_{\widetilde{q}=\widetilde{t}_{1,2};\; \widetilde{b}_{1,2}}
m_{\widetilde{q}}^4
\left(\log\frac{m_{\widetilde{q}}^2}{Q^2}-\frac{3}{2}\right)
-2\sum_{q=t,b}m_q^4
\left(\log\frac{m_q^2}{Q^2}-\frac{3}{2} \right) \right\} ,
\ee
where  we have included the top/stop and bottom/sbottom
contributions, with masses
\be
\begin{array}{rl}
\label{mtopbottom}
m_t^2 & = h_t^2\left|H_2\right|^2  \\
m_b^2 & = h_b^2\left|H_1\right|^2 ,
\end{array}
\ee
and $\mst12^2$, $\msb12^2$
being the two eigenvalues of the
stop and sbottom squared mass matrices
\be
\label{masastop}
M^2_{\st}=\left(
\begin{array}{cc}
\mstl^2 & m^2_{X_t} \\
(m^2_{X_t})^* & \mstr^2
\end{array}
\right)
\ee
\be
\label{masasbottom}
M^2_{\sb}=\left(
\begin{array}{cc}
\msbl^2 & m^2_{X_b} \\
(m^2_{X_b})^* & \msbr^2
\end{array}
\right).
\ee

In Eq.~(\ref{masastop}),
$m_{\;\widetilde{t}_R}^2$ and $m_{\;\widetilde{t}_L}^2$ are the
mass parameters of the left-handed
and right-handed stops, while $m_{X_t}$ denotes the left--right stop
mixing mass parameter. These mass parameters are explicitly given by

\bea
\label{masastlr}
\mstl^2 & = & m_Q^2
+h_t^2\left|H_2\right|^2+\frac{1}{4}(g^2-\frac{1}{3}g'^2)\left( \left|
H_1\right|^2-\left|H_2\right|^2\right) \nonumber \\
\mstr^2 & = & m_U^2+h_t^2\left|H_2\right|^2+\frac{1}{3}g'^2\left( \left|
H_1\right|^2-\left|H_2\right|^2\right) \\
m_{X_t}^2 & = & h_t\left(A_t H_2-\mu H_1^*\right) ,\nonumber
\eea
where the terms depending on the gauge couplings come from the
effective potential D-terms.
Similarly, in Eq.~(\ref{masasbottom}) $\msbl^2$ and
$\msbr^2$ are the mass parameters of left-handed and right-handed sbottoms,
and $m_{X_b}$ denotes the left--right sbottom mixing mass.
They are given by
\bea
\label{masasblr}
\msbl^2 & = & m_Q^2
+h_b^2\left|H_1\right|^2-\frac{1}{4}(g^2+\frac{1}{3}g'^2)\left( \left|
H_1\right|^2-\left|H_2\right|^2\right) \nonumber \\
\msbr^2 & = & m_D^2+h_b^2\left|H_1\right|^2-\frac{1}{6}g'^2\left( \left|
H_1\right|^2-\left|H_2\right|^2\right) \\
m_{X_b}^2 & = & h_b\left(A_b H_1-\mu H_2^*\right). \nonumber
\eea
The mass parameters $m_Q$, $m_U$, $m_D$, $A_t$ and $A_b$,
in (\ref{masastlr}) and (\ref{masasblr}), are the
soft breaking masses for the left-handed stop/sbottom
doublets, right-handed stop and sbottom
singlets, top and bottom trilinear
couplings, respectively, while $\mu$ is the
supersymmetric Higgs mass parameter.

The stop and sbottom mass eigenvalues are given by
\be
\label{masast}
\mst12^2=\frac{\mstl^2+\mstr^2}{2}\pm
\sqrt{ \left(\frac{\mstl^2-\mstr^2}{2}\right)^2
+\left|m_{X_t}^2\right|^2 }
\ee
\be
\label{masasb}
\msb12^2=\frac{\msbl^2+\msbr^2}{2}
\pm\sqrt{ \left(\frac{\msbl^2-\msbr^2}{2}\right)^2
+\left|m_{X_b}^2\right|^2 }
\ee

The content of this paper is as follows:
In section 2 we  re-analyse the
effective potential computation of the neutral Higgs masses, putting
special emphasis in the stop decoupling effects. For simplicity
of presentation,
we shall concentrate there on the case of large values
of the ratio of the two Higgs vacuum expectation values, $\tan\beta$,
and equal soft supersymmetry breaking parameters for the left-
and right-handed stops, ignoring momentarily the bottom--sbottom
contributions. The two-loop leading-log expression for the Higgs
masses will be first presented within this framework. In section 3 we
shall generalize the results of section 2 to the case of
non-degenerate squark mass parameters and arbitrary values of
$\tan\beta$ and the squark mixing mass parameters. The inclusion
of the bottom- and sbottom-dependent corrections will be
discussed. We also define, from the running masses of neutral Higgs
bosons obtained from the effective potential, the
corresponding pole masses by the
inclusion of self-energies where we keep the leading contributions
coming from the top/stop and bottom/sbottom propagation.
We reserve section 4 for our conclusions, and appendices A and B for the
explicit analytical expressions of the different self-energies
and the  gaugino/Higgsino
corrections to the Higgs mass matrix elements,
respectively.

\nsect{The limiting case of large $\tan\beta$ and $m_Q=m_U$}

In this section we  study in detail the simple limiting case of the MSSM
with very large values of $\tan\beta$. This limiting
value is more
than an academic exercise since it provides, for a given set of the
other free parameters of the theory,
the absolute upper bound on the mass of the
lightest CP-even Higgs boson and is therefore useful for practical
applications.  We shall set here the $\mu$-parameter to zero and
neglect the bottom Yukawa
coupling\footnote{This is an excellent approximation
even for moderately large values of $\tan\beta$.} $h_b$.
Furthermore, we shall
assume the case $m_Q=m_U\equiv M$ to simplify
the analysis.  All these constraints
will be relaxed in the next section, where
the general case will be studied.

A good description of the
effective potential is given by just taking $H_1=0$ and $H_2\equiv H$.
For scales $Q>M$ it can be written as
\be
\label{potenciallim}
V=\Lambdatilde+\mtilde2\modh^2+\frac{1}{2}\lambdatilde\modh^4+
V^{(1)}
\ee
where $\Lambdatilde\propto M^4$ is the one-loop vacuum energy induced by
supersymmetry breaking,  $\lambdatilde$ is fixed by supersymmetry to
\be
\label{cuartico}
\lambdatilde = \frac{1}{4}\left(g^2+g'^2\right),
\ee
and  $V^{(1)}$ is the one-loop effective potential (\ref{potgeneral})
where only stop and top
contributions  are taken into account.
In the limiting case we are considering,
and neglecting ${\cal O}(g^4,g^2g'^2,g'^4)$ contributions
in radiative corrections, the masses
contributing to $V^{(1)}$ can be written as:
\be
\begin{array}{rl}
\label{masaslim}
\mst12^2 & =  M^2+\left(h_t^2-\frac{1}{2}\lambdatilde\right)\modh^2
\pm h_t A_t \modh \\
m_t^2 & =  h_t^2\modh^2
\end{array}
\ee
and the  $\beta$-functions of the tilded parameters in (\ref{potenciallim})
are\footnote{We define here $\beta_X\equiv \partial X/\partial\log Q^2$.}
\be
\begin{array}{rl}
\label{betas}
16\pi^2\beta_{\mtilde2} & =  3h_t^2\left(\mtilde2+2M^2+A_t^2\right) \\
16\pi^2\beta_{\; \lambdatilde} & =  0
\end{array}
\ee

For scales $Q<M$ we have to properly take
into account the process of stop decoupling. In a mass-independent
renormalization scheme~\cite{DR} this can be explicitly done as
follows\footnote{Similar approaches can be found in Ref.~\cite{dec}.}.
We expand the $\mst12$ contribution to (\ref{potenciallim}) around
$\log(M^2/Q^2)$ as:
\bea
\label{desarrollo}
& & \frac{3}{32\pi^2}\sum_{\widetilde{q}=
\widetilde{t}_{1,2}}m_{\widetilde{q}}^4
\left(\log\frac{m_{\widetilde{q}}^2}{Q^2}-\frac{3}{2}\right)
=  \frac{3}{16\pi^2} M^4
\left(\log\frac{M^2}{Q^2}-\frac{3}{2}\right) \nonumber \\
& + & \frac{3}{16\pi^2}
\left\{ \left[2M^2\left(h_t^2-\frac{1}{2}\lambdatilde\right)
+h_t^2 A_t^2\right]\log\frac{M^2}{Q^2}-
2M^2 h_t^2\right\}\modh^2 \nonumber \\
& + &  \frac{3}{16\pi^2}
\left\{  h_t^2(h_t^2-\lambdatilde)\log\frac{M^2}{Q^2}
+h_t^2\left[\left(h_t^2-\frac{1}{2}\lambdatilde\right)\frac{A_t^2}{M^2}
-\frac{h_t^2}{12}\frac{A_t^4}{M^4}\right] \right\} \modh^4 \nonumber \\
& + & {\cal O}(\modh^6)
\eea
and absorb the renormalizable terms in (\ref{desarrollo}) in a redefinition
of the low energy parameters:
\bea
\label{redefinicion}
\Lambda & = & \Lambdatilde + \frac{3}{16\pi^2}M^4\left(\log\frac{M^2}{Q^2}
-\frac{3}{2}\right) \nonumber \\
m^2 & = & \mtilde2+
\frac{3}{16\pi^2}\left\{\left[2M^2\left(h_t^2-\frac{1}{2}\lambda
\right)+h_t^2 A_t^2\right]\log\frac{M^2}{Q^2}-2M^2 h_t^2\right\}  \\
\lambda & = & \lambdatilde +
\frac{3}{8\pi^2} \left\{ h_t^2\left(h_t^2-\lambda\right)
\log\frac{M^2}{Q^2}+
h_t^2\left[\left(h_t^2-\frac{1}{2}\lambda\right)\frac{A_t^2}{M^2}
-\frac{1}{12}h_t^2\frac{A_t^4}{M^4}\right]\right\} \nonumber
\eea
where all expressions are expanded up to one loop.

Using now (\ref{betas}) and (\ref{redefinicion}) we obtain
the one-loop $\beta$-functions for
the  untilded parameters
\bea
\label{betasns}
16\pi^2\beta_{\Lambda} & = & 16\pi^2\beta_{\; \Lambdatilde}-3\;
M^4 \nonumber \\
16\pi^2\beta_{m^2} & = &  3(h_t^2 +\lambda)m^2 \\
16\pi^2\beta_{\lambda} & = & -6h_t^2(h_t^2-\lambda) \nonumber
\eea
Notice that (\ref{betasns}) are the  one-loop
$\beta$-functions of the Standard
Model (as they should) to the order
of approximation we are working. In the same
way $\widetilde{H}=H$ and $\widetilde{h}_t=h_t$,
since neither the anomalous
dimension nor the top-quark Yukawa coupling are affected, in our
approximation,   when passing through the threshold scale $M$.

Using now (\ref{potenciallim}), (\ref{desarrollo})
and (\ref{redefinicion}) we can cast
the effective potential for scales $Q<M$ as
\be
\label{potencialsub}
V=\Lambda+m^2\modh^2+\frac{1}{2}\lambda\modh^4
-\frac{3}{16\pi^2}m_t^4\left(\log\frac{m_t^2}{Q^2}-\frac{3}{2}\right)
+G(\modh)
\ee
where $G(\modh)={\cal O}(\modh^6)$, which
contains the higher dimensional terms arising
from (\ref{desarrollo}), is
scale independent and can therefore
be `frozen' at the scale $Q=M$. It is given by
\bea
\label{G}
G & = &  \frac{3}{32\pi^2} \left\{
\sum_{\widetilde{q}=\widetilde{t}_{1,2}}m_{\widetilde{q}}^4
\left(\log\frac{m_{\widetilde{q}}^2}{M^2}-\frac{3}{2}\right) \right. \\
&+& \left. 3M^4 + 4M^2h_t^2 \modh^2-2h_t^2
\left[ \left(h_t^2-\frac{1}{8}(g^2+g'^2)\right)
\frac{A_t^2}{M^2}-\frac{1}{12}h_t^2\frac{A_t^4}{M^4}\right]
\modh^4 \right\} . \nonumber
\eea

The effective potential (\ref{potencialsub}) is a
function of $x\equiv \modh^2 \equiv \phi^2+\psi^2$,
where $\phi\equiv {\rm Re} H$, $\psi\equiv {\rm Im} H$.
Imposing the condition of minimum
at $(\phi,\psi)=(v,0)$,  where we are taking
$v(\mbart)=174.1$ GeV~\cite{CEQR,CEQW},
one obtains a massless scalar (neutral Goldstone boson) along the
direction ${\rm Im}H$. The condition of minimum amounts
to ${\rm d}V(x)/{\rm d}x=0$ and can be written, at $Q^2=\mbart^2$, as
\be
\label{minimo}
m^2=-\lambda v^2 -\frac{3}{8\pi^2}h_t^2 \mbart^2- \frac{{\rm d}G}
{{\rm d}x}
\ee
where we are neglecting two-loop corrections.

Imposing now condition (\ref{minimo}) one gets for the Higgs mass
\be
\label{masahiggs}
m_h^2(\mbart) = 2\lambda_{\rm eff}(\mbart)\;  v^2(\mbart)
\ee
where
\be
\label{lambdaeffec}
\lambda_{\rm eff}=\lambda+ \frac{{\rm d}^2G}{{\rm d}x^2}.
\ee

The usual procedure is to neglect the term
${\rm d}^2G/{\rm d}x^2$ in (\ref{masahiggs}), in which case
the usual expression for the Higgs mass squared in the leading order
approximation, $2\lambda v^2$, is obtained.
We shall now take into account the
contribution from the
$G$-function and see how it modifies the final result. We shall first
analyse the case of zero mixing.

\subsection{The case of zero mixing}

In this case $A_t=0$, and the expression
for $G$ in (\ref{G}) simplifies a lot.
Using (\ref{redefinicion}) and
(\ref{betasns}) one can, to one loop, expand $\lambda$ as
\bea
\label{exp}
\lambda(\mbart)&=&
\lambda(M)-\beta_{\lambda}(M)\log\frac{M^2}{\mbart^2} \nonumber \\
&=& \frac{1}{4}(g^2+g'^2)+\frac{3}{8\pi^2}h_t^2(h_t^2-\lambda)
\log\frac{M^2}{\mbart^2}
\eea
while ${\rm d}^2G/{\rm d}x^2$ is found to give
\be
\label{Gseg}
\frac{{\rm d}^2G}{{\rm d}x^2} = \frac{3}{8\pi^2}
h_t^2(h_t^2-\lambda)\log\frac{M^2+\mbart^2}{M^2}
-\frac{3}{16\pi^2}h_t^4\frac{M_Z^2}{M^2+\mbart^2}
\ee
where all couplings are considered at the scale $M$
(remember that the $G$-term was
frozen at that scale) and we have neglected ${\cal O}(g^4)$ terms.

Putting (\ref{exp}) and (\ref{Gseg}) together, the effective coupling
(\ref{lambdaeffec}) can be written as
\be
\label{efflambda}
\lambda_{\rm eff}(\mbart)=\frac{1}{4}(g^2+g'^2)
+\frac{3}{8\pi^2}h_t^2(h_t^2-\lambda_{\rm eff})
\log\frac{M^2+\mbart^2}{\mbart^2}
-\frac{3}{16\pi^2}h_t^4\frac{M_Z^2}{M^2+\mbart^2}
\ee
where all couplings on the right-hand side are
considered at the scale $M$.

Comparison of (\ref{exp}) with (\ref{efflambda}) shows that the effect of
including the $G$-term in
(\ref{potencialsub}) is, apart from the last (mixed
Yukawa--gauge) term in (\ref{efflambda}) which is numerically unimportant,
to resum the  series
$$ \sum_{n=1}^{\infty}\frac{(-1)^{n+1}}{n}
\left(\frac{\mbart^{2n}}{M^{2n}}\right)$$
from the expansion parameter $\log(M^2/\mbart^2)$ to the physical one
$\log((M^2+\mbart^2)/\mbart^2)$,
which goes to zero in the supersymmetric limit
$M\rightarrow 0$, and makes the radiative
corrections vanish in that limit.

In other words the resummation
from the $G$-function allows us to replace the
expansion (\ref{exp}) by the expansion
\be
\label{expMs}
\lambda_{\rm eff}(\mbart)=
\lambda_{\rm eff}(M_S)-\beta_{\lambda_{eff}}(M_S)t
\ee
where the expansion parameter is
\be
\label{t}
t=\log\frac{M_S^2}{\mbart^2}
\ee
and
\be
\label{MS}
M_S^2=M^2+\mbart^2 .
\ee

\subsection{The case $A_t\neq 0$}

Using again (\ref{redefinicion}) and (\ref{betasns}) one can write
\bea
\label{expmix}
\lambda(\mbart) & = & \lambda(M)-
\beta_{\lambda}(M)\log\frac{M^2}{\mbart^2}\nonumber \\
&=&  \frac{1}{4}(g^2+g'^2)+\frac{3}{8\pi^2}h_t^2(h_t^2-\lambda)
\log\frac{M^2}{\mbart^2} \\
&+ &
\frac{3}{8\pi^2} h_t^2
\left[\left(h_t^2-\frac{1}{8}(g^2+g'^2)\right)\frac{A_t^2}{M^2}
-\frac{1}{12}h_t^2\frac{A_t^4}{M^4}\right] \nonumber
\eea
In the absence of the $G$-contribution in (\ref{lambdaeffec}),
Eq.~(\ref{expmix})
gives the usual one-loop leading-log
contribution to the Higgs mass. Introducing
now the $G$-function, one can write (\ref{lambdaeffec}) as:
\be
\label{efflamix}
 \lambda_{\rm eff}(\mbart)
 =  \frac{1}{4}(g^2+g'^2)+\frac{3}{8\pi^2}h_t^2
(h_t^2-\lambda_{\rm eff})
\log\frac{M_S^2}{\mbart^2} + \Delta_{\rm th}\lambda_{\rm eff} ,
\ee
where the `threshold' contribution to the coupling is given by
\bea
\label{umbral}
\Delta_{\rm th}\lambda_{\rm eff} & = &
\frac{3}{64 v^4\pi^2}\left\{\left(2m_t^2-\frac{1}{2}M_Z^2+m_t A_t\right)^2
\log\left(1+\frac{m_tA_t-\frac{1}{4}M_Z^2}{M_S^2}\right)
\right. \nonumber \\
&+ & \left(2m_t^2-\frac{1}{2}M_Z^2-m_t A_t\right)^2
\log\left(1-\frac{m_tA_t+\frac{1}{4}M_Z^2}{M_S^2}\right)
 \\
&- & m_t A_t \left\{
\left(M_S^2-\frac{1}{4}M_Z^2+m_t A_t \right)
\left[\log\left(1+
\frac{m_t A_t-\frac{1}{4}M_Z^2}{M_S^2}\right)-1\right]\right.
\nonumber \\
& - & \left.\left. \left(M_S^2-\frac{1}{4}M_Z^2-m_t A_t \right)
\left[\log\left(1-
\frac{m_t A_t+\frac{1}{4}M_Z^2}{M_S^2}\right)-1\right]\right\}
\right\} \nonumber
\eea
and it can be easily checked that
\be
\label{expumbral}
\Delta_{\rm th}\lambda_{\rm eff}=
\frac{3}{8\pi^2} h_t^2\left[\left(h_t^2-\frac{1}{8}(g^2+g'^2)\right)
\frac{A_t^2}{M_S^2}
-\frac{1}{12}h_t^2\left(\frac{A_t^2}{M_S^2}\right)^2\right]
+\cdots
\ee
where the ellipsis denotes the contribution from higher-dimensional
terms in $m_t^2/M_S^2$ and $A_t^2/M_S^2$.

We can see that the effect of taking into account the $G$-function in
(\ref{potencialsub}) is twofold:
\begin{itemize}
\item
It makes a resummation from $\log(M^2/\mbart^2)$
to the physical expansion parameter
$\log(M_S^2/\mbart^2)$ in the one-loop radiative correction,
as well as resummation from $A_t^2/M^2$ to $A_t^2/M_S^2$ in the
threshold term of (\ref{expmix}).
\item
It generalizes the threshold correction of Eq.~(\ref{expmix})
to include higher order effects
in powers of $A_t^2/M_S^2$ in Eq.~(\ref{umbral}).
\end{itemize}
Therefore the net effect of the $G$-function is to change the expansion
(\ref{expmix}) into
\be
\label{expMsmix}
\lambda_{\rm eff}(\mbart)=\lambda_{\rm eff}(M_S)-
\beta_{\lambda_{\rm eff}}(M_S)t
\ee
where
\be
\label{inicial}
\lambda_{\rm eff}(M_S)=
\frac{1}{4}(g^2+g'^2)+\Delta_{\rm th}\lambda_{\rm eff} .
\ee

\subsection{Two-loop leading-log expansion}

As we have seen in Ref.~\cite{CEQW} one can obtain the two-loop leading-log
correction by expanding the parameter $\lambda$
to order $(\log(M^2/\mbart^2))^2$, as
\be
\label{expmix2}
\lambda(\mbart)=\lambda(M)-\beta_{\lambda}(M)\log\frac{M^2}{\mbart^2}
+\frac{1}{2}\beta'_{\lambda}\left(\log\frac{M^2}{\mbart^2}\right)^2,
\ee
where, since we are now considering two-loop corrections, we have
evaluated the quartic coupling at the on-shell top quark mass scale
and the prime denotes derivative with respect to $\log(Q^2)$.
In the previous subsection we have seen that
the effect of including the $G$-function
in the effective potential is to resum $M^2$
to $M_S^2$ in the first two terms of
(\ref{expmix2}) and  replace the threshold correction in the first one by
$\Delta_{\rm th}\leff$ in (\ref{umbral}) (in Ref.~\cite{CEQW},
this resummation effect was assumed to be true to assure a proper
behaviour of the radiative corrections for $M = 0$).
To prove resummation in the last term of
(\ref{expmix2}) one would need to use the whole
two-loop effective potential in the
MSSM. In the absence of such a calculation
we shall assume that this happens, since we
have already proved that $t$ in (\ref{t}) is the
physical expansion parameter in the
one-loop calculation, and in addition the
numerical relevance of the resummation in the
two-loop corrections is expected to be tiny. We shall therefore
consider as a starting point
\bea
\label{efflamix2}
\leff(\mbart)& = & \leff(M_S)-\beta_{\leff}(M_S)t
+\frac{1}{2}\beta'_{\leff}(\mbart)t^2 +\cdots \nonumber \\
& = & \leff(M_S)-\beta_{\leff}(\mbart)t
-\frac{1}{2}\beta'_{\leff}(\mbart)t^2+\cdots
\eea
where $\leff(M_S)$ is given in (\ref{inicial})
and $t$ and $M_S$ are obtained from Eqs.~(\ref{t}) and
(\ref{MS}).
Defining $\beta_{\leff} = a_{\leff} \leff + b_{\leff}$,
it follows that
\be
\lambda_{\rm eff}(\mbart) = \lambda_{\rm eff}(M_S)  \left( 1
- a_{\leff}(\mbart) \; t \right) - b_{\leff}(\mbart) \; t \;
\left( 1
- a_{\leff}(\mbart) \; t \right) -\frac{1}{2}\beta'_{\leff}(\mbart)t^2.
\ee
{}From the renormalization group equations of the quartic coupling,
it follows that the
term $1 - a_{\leff}(\mbart) \;t = \xi^{-4}(\mbart)$, where
$\xi$, the Higgs field anomalous dimension, is:
\be
\label{anomdim}
\xi(\mbart)=1+\frac{3}{32\pi^2}h_t^2(\mbart)\;t.
\ee
Recalling that $v^2(M_S) = v^2(\mbart) \; \xi^{-2}(\mbart)$, from
Eq.~(\ref{masahiggs}) we get
\be
\label{decmasa}
m_h^2(\mbart) = m_h^2(M_S)\;\xi^{-2}(\mbart)+
\Delta_{\rm rad\; }m_h^2(\mbart).
\ee
In the above,
\be
\label{masafroz}
m_h^2(M_S) = 2 \lambda_{\rm eff}(M_S)\;v^2(M_S),
\ee
$\lambda_{\rm eff}(M_S)$ is given in Eq.~(\ref{inicial})
with all couplings and masses
evaluated at the scale $M_S$,
and $\Delta_{\rm rad\; }m_h^2(\mbart)$ is given by
\bea
\label{radmas}
\Delta_{\rm rad\; }m_h^2(\mbart)& = &
2v^2(\mbart)\left[
- b_{\leff}(\mbart) \; t \;
\left( 1
- a_{\leff}(\mbart) \; t \right) -\frac{1}{2}\beta'_{\leff}(\mbart)t^2
\right] \nonumber \\
& = & \frac{3}{4\pi^2}\frac{\mbart^4}{v^2(\mbart)}\;t
\left[1+\frac{1}{16\pi^2}\left(\frac{3}{2}h_t^2-32\pi\alpha_3\right)t
\right]
\eea
where all couplings in (\ref{radmas}) are evaluated at
the scale $Q^2=\mbart^2$,
$\alpha_3$ is the strong gauge coupling, and we have used
$\beta_{h_t}$ in the evaluation of (\ref{radmas})~\cite{CEQW}.
Observe that, if we replace $\Delta_{\rm th}\lambda_{\rm eff}$
by its expansion in powers of $A_t/M_S$, Eq.~(\ref{expumbral}),
we neglect the small terms depending on the weak gauge couplings,
and we re-express the values of the couplings at $M_S$ by their
expressions at $\mbart$ using the appropriate $\gamma$- and
$\beta$-functions, we obtain
\begin{eqnarray}
m_h^2& = & M_Z^2\left( 1-\frac{3}{8\pi^2}\frac{\mbart^2}
{v^2}\ t\right) \nonumber \\
\label{mhsm}
& + & \frac{3}{4\pi^2}\frac{\mbart^4}{v^2}\left[ \frac{1}{2}\tilde{X}_t + t
+\frac{1}{16\pi^2}\left(\frac{3}{2}\frac{\mbart^2}{v^2}-32\pi\alpha_3
\right)\left(\tilde{X}_t t+t^2\right) \right]
\end{eqnarray}
where
\begin{eqnarray}
\label{stopmix}
\tilde{X}_{t} & =& \frac{2 A_t^2}{M_{S}^2}
                  \left(1 - \frac{A_t^2}{12 M_{S}^2} \right).
\end{eqnarray}
Equation (\ref{mhsm}) is equivalent to
Eq.~(9) of Ref.~\cite{CEQW} in the large $\tan\beta$ regime.

Although we have used, as simplifying hypothesis,
the case where $\tan\beta\gg 1$ and $\mu = 0$, all
the results are also valid for
the case $m_A\sim M_S$ and any value of $\tan\beta$
and $\mu$. The only
change in the final results is that
\bea
\label{relacion}
(g^2+g'^2) &  \longrightarrow & (g^2+g'^2)\cos^2 2\beta \nonumber \\
A_t & \longrightarrow & A_t-\mu\cot\beta
\eea
in $\lambda(M)$ and $\leff(M_S)$.

Finally, we want to conclude this section with a comment about
the physical interpretation of the decomposition (\ref{decmasa}).
The term $m_h^2(M_S)$  comes from the
scale independent  part of the MSSM effective
potential frozen at the scale $M_S$, where we have already subtracted the
contribution
evolving with $\log(M_S^2/\mbart^2)$.
Since this term is scale independent, it is
evolved to the scale
$\mbart$ with the corresponding power of the anomalous dimension
of the Higgs field.
On the contrary, the term $\Delta_{\rm rad\; }m_h^2$, which arises
from renormalizable terms after the stops
are decoupled at the high scale,  is computed at
the low scale $\mbart$. Had we considered the whole MSSM effective potential
at the scale $\mbart$~\cite{ERZ}, we would have been neglecting the stop
decoupling at the scale $M_S$.
This is shown in Fig.~1 (dotted lines) where we
plot $m_h$ as a function of $M_S$ for
a pole top-quark mass
$M_t=175$ GeV,
vanishing mixing, $A_t=\mu=0$, and large ($\tan\beta=15$) and small
[infrared (IR) fixed point solution:
$\sin\beta\sim (200\; {\rm GeV}/M_t)$] values of $\tan\beta$.
Had we evolved the Higgs mass obtained from
the whole MSSM effective potential
(including the logarithmic terms) at $M_S$, to the scale
$\mbart$ with the anomalous dimension factor $\xi^{-2}(\mbart)$,
we would have made an error associated with the
non-exact scale invariance of the effective
potential in the one-loop approximation,
as was observed in~\cite{CEQR}.  This is shown in the
dashed lines of Fig.~1
and was also noticed in Ref. \cite{CEQW}.  In fact the latter procedure
would lead to an expression of $\Delta_{\rm rad\; }m_h^2$,
where the second term inside the
square brackets has an extra factor of 2
[see Eq.~(\ref{mhsm})].   The solid
lines in Fig.~1 show the corresponding
value for the Higgs mass, while considering
 Eq.~(\ref{decmasa}) with the whole expression for
$\Delta_{\rm th} \lambda_{\leff}$
from Eq.~(\ref{umbral}).

The dependence of (\ref{decmasa}) on the mixing is shown in Fig.~2
(solid lines) where we plot $m_h$ as a function of $A_t$, for $\mu=0$,
the same values of $\tan\beta$ as in Fig.~1 and
$M_S=1$~TeV. For comparison we also plot
(dashed lines) the
corresponding mass, using the approximate expression for the
threshold contribution to the Higgs quartic couplings,
Eq.~(\ref{mhsm}).  We see
that this approximation, which was used in~\cite{CEQW}, is very
good up to the maximum of the
curve, which means that the absolute upper bound on
the mass of the lightest Higgs boson
previously obtained remains unchanged for any
value of the mixing. Of course for very large
values of the mixing there is a departure between the two curves.
We have deliberately omitted from Figs.~1 and 2 the small D-term
contributions in Eq.~(\ref{umbral}) to compare with
previous results which did not make use of them \cite{ERZ,CEQW}.
They will be included in the general analysis of section 3.

\nsect{The general case}

Having understood the simplified case explained in section 2, we
can proceed with the general case. Let us first focus on the
behaviour of the renormalized Higgs quartic couplings for low
values of the CP-odd Higgs mass and values of the
mass parameter $m_Q$ different  from $m_U$.
In this case, the effective potential at scales above
${\rm Max}(m_Q,m_U,m_D)$ is
obtained through the general expression, Eq.~(\ref{potgeneral}).
Expanding
the effective potential in powers of the Higgs fields, it is
easy to identify the form of the effective quartic couplings
at the scale $m_Q$, $\lambda_i(m_Q)$, in a way completely analogous
to what we have done for the case $m_Q = m_U$ in section 2.
Assuming for definiteness that $m_Q > m_U > m_D$,
\be
\lambda_i(m_Q) = \lambda_i(m_U) +
\beta_{i}^{QU}(m_U) \; \widetilde{t}_{QU}
+ \left(\beta_{i}^{QU}\right)' \; \frac{\;\widetilde{t}_{QU}^{\; 2}}{2}
\ee
[see e.g. the second equality in Eq.~(\ref{efflamix2})]
where $\widetilde{t}_{QU} = \log(m_Q^2/m_U^2)$, $\beta_i$
denotes the $\beta$-function of the Higgs quartic couplings,
$\beta_i = d \lambda_i/d\log(Q^2)$ and the superscript $QU$
denotes its expression at
energy scales between $m_Q$ and $m_U$. In general,
we can write,
\be
\beta_i = a_i \lambda_i + b_i, \;\;\;\;\;\;\;\;\;
\beta_i' \simeq a_i b_i + b'_i ,
\ee
where we have only kept the dominant, Yukawa-coupling-dependent
contributions to $\beta'_i$. We omit the scale dependence of
$\beta_i'$ since it is a higher-order effect.
Hence,
\be
\lambda_i(m_U) = \lambda_i(m_Q) \left( 1 - a_i \;
\widetilde{t}_{QU} \right) - b^{QU}_i(m_U)
\; \widetilde{t}_{QU} \left( 1 -
a_i \; \widetilde{t}_{QU} \right) -
\left(\beta_{i}^{QU}\right)' \; \frac{\;\widetilde{t}_{QU}^{\; 2}}{2}.
\ee

The coefficients $a_i$ are linear combinations of the
anomalous dimensions of the Higgs fields $H_i$, which are independent
of the squark fields.
The same procedure as used above may be used to connect the value of
the quartic couplings at $m_D$ with their values at $m_U$,
and finally the values at $m_D$ with their values at $\mbart$.
Keeping only the dominant terms, we obtain

\begin{eqnarray}
\lambda_i(\mbart) & = &\lambda_i(m_Q) \left( 1
- a_i(\mbart) \; \widetilde{t}_Q \right)
- \left(\beta_{i}^{QU}\right)' \frac{\;\widetilde{t}_{QU}}{2}
\left(\widetilde{t}_{Q} + \widetilde{t}_{U}\right)
\nonumber\\
&-&   b^{QU}_i(\mbart) \; \widetilde{t}_{QU} \; \left[ 1 -
a_i(\mbart) \left(\widetilde{t}_{Q} + \; \widetilde{t}_{U}
\right) \right]
\nonumber\\
& - &  b^{UD}_i(\mbart) \; \widetilde{t}_{UD} \left[ 1 -
a_i(\mbart) \; \left(\widetilde{t}_{U} +
\widetilde{t}_D \right) \right]
- \left(\beta_{i}^{UD}\right)' \; \frac{\;\widetilde{t}_{UD}}{2}
\left(\widetilde{t}_{U} + \widetilde{t}_D \right)
\nonumber\\
& - & b_i(\mbart) \; \widetilde{t}_{D} \left( 1 -
a_i (\mbart)\; \widetilde{t}_{D} \right)
-\beta_{i}' \; \frac{\;\widetilde{t}_{D}^{\; 2}}{2},
\label{lambimt}
\end{eqnarray}
where $\beta_i$ without any superscript
denotes the $\beta$-function values once
the two stops and the two
sbottoms are decoupled, the superscript $XY$, with $X,Y = Q,U,D$,
denotes the functional
form of the $\beta$-functions  at scales between $m_{X}$ and
$m_{Y}$, $\widetilde{t}_{XY} = \widetilde{t}_X - \widetilde{t}_Y$
and $\widetilde{t}_X = \log(m_X^2/\mbart^2)$.
Similar expressions are obtained for a
different hierarchy of the squark mass parameters, with the
only difference that, in the case $m_U > m_Q$ and/or
$m_D > m_Q$, stops and sbottoms should be decoupled at
different scales. For simplicity of presentation
we shall first discuss the
result for the Higgs mass matrix elements in the case under
study and we shall present below
the result in the most general case.

The contribution of
the quartic couplings to the Higgs mass matrix elements is
then given by
\begin{eqnarray}
  M^2_{12} &=&  2 v^2 [\sin \beta \cos \beta (\lambda_3 + \lambda_4) +
     \lambda_6 \cos^2 \beta + \lambda_7 \sin^2 \beta ] - m_A^2
 \sin\beta \cos\beta
\nonumber\\
     M^2_{11} &=& 2 v^2 [\lambda_1  \cos^2 \beta + 2
     \lambda_6  \cos \beta \sin \beta
     + \lambda_5 \sin^2 \beta] + m_A^2 \sin^2\beta
\label{mijl}\\
     M^2_{22} &=&  2 v^2 [\lambda_2 \sin^2 \beta +2 \lambda_7  \cos \beta
     \sin \beta + \lambda_5 \cos^2 \beta] + m_A^2 \cos^2\beta,
\nonumber
\end{eqnarray}
where all terms should be evaluated at the same scale and
we have also included the dependence on the
CP-odd Higgs mass~\cite{HH}. In the above, $v_1 = v \cos\beta$
and $v_2 = v \sin\beta$ are the $H_1$ and $H_2$ vacuum expectation
values, respectively.

Equation (\ref{lambimt})  has a clear interpretation:
The factor $\lambda_i(m_Q)$ in the first term
contains the tree level terms
and all finite contributions to the quartic
couplings, arising  from
the existence of squark mixing and the fact that
$m_D \neq m_Q \neq m_U$.
The factor involving  $a_i$ in the first term contains exactly
the terms necessary to rescale the Higgs mass
matrix elements by the appropriate anomalous dimension factors
from the scale $m_Q$ to the scale $\mbart$, together with the ones
necessary to re-express the
vacuum expectation values of the Higgs fields
appearing in Eq.~(\ref{mijl}) at the scale $\mbart$,
in terms of their
values at the scale $m_Q$ (for the complete expression of the
$\beta$-functions of the Higgs quartic couplings,
see for example Ref.~\cite{HH}). For instance, singling out this
contribution
of the $\lambda_2$ coupling to the matrix element $M_{22}^2$, which
we shall denote by $K_{22}$, we obtain
\begin{eqnarray}
K_{22}(\mbart) &=&2 \; \lambda_2(m_Q)
\left[ 1 - a_2 \log\left( \frac{m_Q^2}{\mbart^2}
\right) \right] \; v_2^2(\mbart)
\nonumber\\
&=& 2 \; \lambda_2(m_Q) \; \xi_2^{-4}(\mbart) \; v_2^2(\mbart)
\\
& & \nonumber \\
&=& 2 \; \lambda_2(m_Q) \; v_2^2(m_Q) \; \xi_2^{-2}(\mbart)\nonumber,
\end{eqnarray}
where $\xi_2$ is the anomalous dimension factor of the Higgs
fields $H_2$.
The contribution of the other couplings to the Higgs mass matrix
elements present similar properties.
Thus, both the tree level term and all finite terms leading
to the non-trivial matching of the quartic couplings at the
scale $m_Q$ may be treated in the same way as the terms proceeding
from the higher-dimensional operator contributions to the
Higgs mass matrix elements. That is, they may be frozen
at the scale $m_Q$ and rescaled with the appropriate anomalous
dimension factors to obtain their expressions at low energies.
This generalizes the result obtained in section 2 for the case
of degenerate squark masses, $m_Q = m_U$.

The following terms in Eq.~(\ref{lambimt}) are the ones which
would be obtained even in the presence of trivial matching
conditions for the quartic Higgs couplings and, as has been
clearly explained in the previous section,  are associated
with the scale-dependent contributions to the effective potential.
The dominant leading-log
contribution to the quartic couplings proceeds from the terms
in the $\beta$-function proportional to the fourth power of
the top and bottom quark Yukawa couplings, which are given
by
\be
\lambda_i(\mbart) - \lambda_i(m_Q) = - \frac{b_i^Y}{2}
\left( \log\left(\frac{m_Q^2}{\mbart^2}\right)
+ \log\left(\frac{m_X^2}{\mbart^2}\right) \right),
\label{llog}
\ee
where $b_1^Y = -3 h_b^4/8 \pi^2$ and $b_2^Y = -3 h_t^4/8 \pi^2$,
$h_b$ and $h_t$ are the bottom and top quark Yukawa
couplings, and $X = D, U$ for $i = 1,2$,
respectively. Although $b_3$ and
$b_4$ also present a quartic dependence on the Yukawa couplings,
such dependence is absent from $b_3 + b_4$.
As may be easily proved using Eq.~(\ref{mijl}), and
following the same procedure
as in section 2, in the case of no mixing
the contribution of the
higher-order operator to the \lq 22' (\lq 11')
Higgs mass matrix elements
allows us to replace the factors $m_Q^2$ and $m_U^2$
($m_Q^2$ and $m_D^2$)
in the leading order expressions by $m_{Q}^2
+ \mbart^2$ and $m_U^2 + \mbart^2$
($m_Q^2 + m_b^2$ and $m_D^2 + m_b^2$).
The same occurs with
the D-term contributions proportional to the square of
products of the weak couplings and the Yukawa couplings.
It is hence convenient to define
\begin{eqnarray}
t_Q^U & = & \log\left(\frac{m_Q^2 + \mbart^2}{\mbart^2}\right) \;\;\;\;\;
{\rm and} \;\;\;\;
t_U = \log\left(\frac{m_U^2 + \mbart^2}{\mbart^2}\right),
\nonumber\\
t_Q^D & = & \log\left(\frac{m_Q^2 + m_b^2}{m_b^2}\right) \;\;\;\;\;
{\rm and} \;\;\;\;
t_D  =  \log\left(\frac{m_D^2 + m_b^2}{m_b^2}\right) \;\;\;\;\;
\label{tqtu}
\end{eqnarray}
while $t_{QU} = t_Q^U - t_U$ and $t_{QD} = t_Q^D - t_D$.
Observe that in the denominators of the expressions for
$t_D$ and $t_Q^D$ we have written $m_b$ instead of $\mbart$
since we know that, upon consideration of the whole
supersymmetric contributions to the physical masses,
radiative corrections should
vanish in the supersymmetric limit; we should thus use
expansion parameters $t_Q^D$ and $t_D$, which vanish
in the limit $m_Q,\; m_D\rightarrow 0$. This
change has negligible effects on the Higgs mass computation.

In the general case,
the way to proceed to obtain the Higgs mass matrix elements
at the scale $\mbart$ is the following:  the
CP-even Higgs mass matrix elements may be decomposed
in three terms, namely:
\begin{eqnarray}
{\cal{M}}^2_{ij}(\mbart) & = &
\overline{M}^2_{ij}(\mbart)
+ \left(\widetilde{{\cal{M}}}^2_{ij}
(M_{st}) \right)_{\widetilde{t}}
\left(\xi_{i}^{\widetilde{t}}(\mbart)
\xi_j^{\widetilde{t}}(\mbart)\right)^{-1}
\nonumber\\
& + &
\left(\widetilde{{\cal{M}}}^2_{ij}(M_{sb}) \right)_{\widetilde{b}}
\left(\xi_{i}^{\widetilde{b}}(\mbart)
\xi_j^{\widetilde{b}}(\mbart)\right)^{-1},
\label{m2ijmt}
\end{eqnarray}
where  $\xi_{i}(\mbart)$ denote
the anomalous dimension factors,
\bea
\xi_1^{\widetilde{b}}(\mbart) =
1 + \frac{3 h_b^2}{32 \pi^2} \; t^{\widetilde{b}}_1 \;\;\;\; &&
\;\;\;\;\;\;\;
\xi_2^{\widetilde{b}} (\mbart)= 1 + \frac{3 h_t^2}{32 \pi^2} \;
t^{\widetilde{b}}_1
\nonumber\\
\xi_1^{\widetilde{t}}(\mbart) =
1 + \frac{3 h_b^2}{32 \pi^2} \; t^{\widetilde{t}}_1 \;\;\;\; &&
\;\;\;\;\;\;\;
\xi_2^{\widetilde{t}} (\mbart)= 1 + \frac{3 h_t^2}{32 \pi^2} \;
t^{\widetilde{t}}_1
\label{anomdij}
\eea
where for convenience we define,
\begin{equation}
t^{\widetilde{b}}_1 = {\rm Max}(t_Q^D,t_D) \;\;\;\;\;\;\;\;\;
t^{\widetilde{t}}_1 = {\rm Max}(t_Q^U,t_U),
\end{equation}
and also, for later use,
\begin{equation}
t^{\widetilde{b}}_2 = {\rm Min}(t_Q^D,t_D)
\;\;\;\;\;\;\;\;\;
t^{\widetilde{t}}_2 = {\rm Min}(t_Q^U,t_U).
\label{tbtt}
\end{equation}
In the above,
$\left(\widetilde{{\cal{M}}}^2_{ij}(M_{st})\right)_{\widetilde{t}}$
$\left(\left(\widetilde{{\cal{M}}}^2_{ij}(M_{sb})
\right)_{\widetilde{b}}\right)$
is the contribution to the
mass matrix elements coming from the terms
frozen at the scale $M_{st}$ ($M_{sb}$),
where the stops (sbottoms) are decoupled,
with $M_{st}^2 = {\rm Max}(m_Q^2 +\mbart^2,m_U^2+\mbart^2)$
$\left(M_{sb}^2 = {\rm Max}(m_Q^2 + m_b^2,m_D^2+ m_b^2)\right)$,
and $\overline{M}^2_{ij}$ are obtained from the mass matrix
elements $M^2_{ij}$, Eq.~(\ref{mijl}), by considering the
one- and two-loop leading logarithm
contributions in the
renormalizable Higgs quartic coupling expressions.
For example, in the case $m_Q > m_U > m_D$, these
contributions to the quartic couplings $\overline{\lambda}_i$
are given by
\be
\overline{\lambda}_i = \lambda_i -
\left(\lambda_i(m_Q) - \lambda_i^{{\rm tree}}(m_Q) \right)
\left( 1 - a_i(\mbart) \; \widetilde{t}_{Q}  \right) ,
\ee
where the $\lambda_i$ are given in Eq.~(\ref{lambimt}),
$\lambda_1^{{\rm tree}} = \lambda_2^{{\rm tree}} =
-\left(\lambda_3^{{\rm tree}}+\lambda_4^{{\rm tree}}\right) =
\left(g^2 + g'^2\right)/4$,
and,
in order to simplify the presentation, we include all
D-term contributions in the definition of the quartic couplings
$\overline{\lambda}_i$.
Observe that, since $\overline{M}_{ij}^2(\mbart)$
contains only the one- and two-loop
leading logarithm expressions independent of the mixing mass
terms, the quartic couplings $\lambda_5$, $\lambda_6$ and
$\lambda_7$ give no contribution to
$\overline{M}_{ij}^2(\mbart)$. Replacing the quartic coupling
$\beta$-functions by their dominant Yukawa coupling dependence,
we obtain, in the general case
\begin{eqnarray}
 \overline{\lambda}_1 &=&
\frac{g^2 + g'^2}{4} \left[ 1 - \frac{3}{8 \pi^2} h_b^2
\;t^{\widetilde{b}}_1 \right]
\nonumber\\
 & + &  \frac{3}{16 \pi^2}\; h_b^4\;
\left(t^{\widetilde{b}}_1 - t^{\widetilde{b}}_2 \right)\; \left[
         1
        + \frac{1}{16 \pi^2}
        \left( \frac{3}{2} \;h_b^2 + \frac{1}{2}\;h_t^2
     - 8\; g_3^2 \right) \left(t^{\widetilde{b}}_1 +
t^{\widetilde{b}}_2  \right) \right]
\nonumber\\
    & + &
           \frac{3}{8 \pi^2}\; h_b^4\;
    t^{\widetilde{b}}_2 \left[
         1 + \frac{1}{16 \pi^2}
        \left( \frac{3}{2} \;h_b^2 + \frac{1}{2}\;h_t^2
     - 8\; g_3^2 \right)
    t^{\widetilde{b}}_2 \right] + \Delta_1^D,
\label{lambda1}
\end{eqnarray}

\begin{eqnarray}
      \overline{\lambda}_2 &=&
\frac{g^2 + g'^2}{4} \left[ 1 - \frac{3}{8 \pi^2} h_t^2
\; t^{\widetilde{t}}_1 \right]
\nonumber\\
 & + &    \frac{3}{16 \pi^2}\; h_t^4\;
\left(t^{\widetilde{t}}_1 - t^{\widetilde{t}}_2 \right) \;
\left[         1
         + \frac{1}{16 \pi^2}
        \left( \frac{3 \;h_t^2}{2} + \frac{h_b^2}{2}
     - 8\; g_3^2 \right)\left(
t^{\widetilde{t}}_1 + t^{\widetilde{t}}_2 \right) \right]
\nonumber\\
      & + &
         \frac{3}{8 \pi^2}\; h_t^4\;t^{\widetilde{t}}_2 \; \left[
         1 + \frac{1}{16 \pi^2}
        \left( \frac{3 \;h_t^2}{2} + \frac{h_b^2}{2}
     - 8\; g_3^2 \right) t^{\widetilde{t}}_2 \right] + \Delta_2^D,
\label{lambda2}
\end{eqnarray}

\begin{eqnarray}
\overline{\lambda}_3 + \overline{\lambda}_4 & = & \Delta_3^D + \Delta_4^D
\nonumber\\
& - & \frac{g^2 + g'^2}{4} \left[ 1 -
\frac{3}{16 \pi^2} h_b^2
\; t^{\widetilde{b}}_1 -
\frac{3}{16 \pi^2} h_t^2
\; t^{\widetilde{t}}_1 \right],
\label{lambda4}
\end{eqnarray}
where we have already performed the top and bottom mass resummations,
leading to the change $\widetilde{t}_X \rightarrow t_X$, with
$X = Q,U,D$. In the above, $h_t$ and $h_b$ denote the top and bottom
quark Yukawa couplings at the scale $\mbart$ and $g_3$ is
the strong gauge coupling
at the same scale. The $\Delta_i^D$ terms are
the additional
leading-log D-term contributions appearing through
the $\beta$-functions of the quartic couplings, which contain
additional terms proportional to the square of the product of
weak gauge couplings and Yukawa couplings.
Their two-loop leading-log contributions are
very small and can be ignored.  Interestingly enough,
once these terms are considered together with the terms
written explicitly  in Eqs.
(\ref{lambda1})--(\ref{lambda4}),
one recovers the expressions for the D-terms
first found in Ref.~\cite{B}.
Defining $\Delta_3^D =
\left( \Delta_3^D \right)_U + \left( \Delta_3^D \right)_D$,
for $m_Q \geq m_D$, we obtain
\begin{eqnarray}
\Delta_1^D & =& \frac{1}{16 \pi^2} g'^2 h_b^2 \; t_{QD}
\nonumber\\
\left(\Delta_3^D\right)_D
& =& - \frac{1}{32 \pi^2} g'^2 h_b^2 \; t_{QD}
\end{eqnarray}
while for $m_D \geq m_Q$ we get
\begin{eqnarray}
\Delta_1^D & =& -\frac{3}{32 \pi^2} \left(\frac{g'^2}{3} +
g^2\right)
h_b^2 \; t_{QD}
\nonumber\\
\left(\Delta_3^D\right)_D
& =& - \frac{3}{64 \pi^2} \left( g^2 - \frac{g'^2}{3} \right)
h_b^2 \; t_{QD}
\\
\Delta_4^D & = &  \frac{3}{32 \pi^2} g^2 h_b^2 \; t_{QD} .
\nonumber
\end{eqnarray}

Analogously, for $m_Q \geq m_U$ we obtain
\begin{eqnarray}
\Delta_2^D & =& \frac{1}{8 \pi^2} g'^2 h_t^2 \; t_{QU}
\nonumber\\
\left(\Delta_3^D \right)_U
& =& - \frac{1}{16 \pi^2} g'^2 h_t^2 \; t_{QU}
\end{eqnarray}
while for $m_U \geq m_Q$ we get
\begin{eqnarray}
\Delta_2^D & =&- \frac{3}{32 \pi^2} \left(-\frac{g'^2}{3} +
g^2\right)
h_t^2 \; t_{QU}
\nonumber\\
\left(\Delta_3^D\right)_U
& =& -\frac{3}{64 \pi^2} \left( g^2 + \frac{g'^2}{3} \right)
h_t^2 \; t_{QU}
\\
\Delta_4^D & = &  \frac{3}{32 \pi^2} g^2 h_t^2 \; t_{QU} .
\nonumber
\end{eqnarray}

Finally notice that $\tan\beta$ is fixed at the scale $m_A$, for
$m_A\leq \mbart$, while for $m_A\geq\mbart$, $\tan\beta$
is given by
\be
\label{tanbeta}
\tan\beta(\mbart)=\tan\beta(m_A)\left[
1+\frac{3}{32\pi^2}(h_t^2-h_b^2)\log
\frac{m_A^2}{\mbart^2} \right] .
\ee
For the case in which the CP-odd Higgs mass $m_A$ is
lower than $M_S = {\rm Max} \left(M_{st},M_{sb}\right)$,
we should have decoupled (strictly speaking)
the heavy Higgs doublet at the scale $m_A$,
and defined an effective quartic coupling
for the light Higgs as $\lambda(m_A)=m_h(m_A)/2v^2$, the low
energy value of it being obtained by the running of the Standard
Model renormalization group equations from the scale $m_A$ to
$\overline{m}_t$ \cite{CEQW}.
For simplicity we have ignored, in our analytical
approximation, the effect of the heavy Higgs doublet decoupling at the
intermediate scale. We partially compensate this effect by relating the
value of $\tan\beta$ at the scale $\overline{m}_t$ with its corresponding
value at the scale $m_A$ through its renormalization-group running,
Eq.~(\ref{tanbeta}).

The Higgs mass matrix elements at the scales $M_{st},M_{sb}$
may be inferred
from the second derivative
of the one-loop effective potential, Eq.~(\ref{potgeneral}), at the
minimum values for the Higgs fields. Consequently, they
can be  obtained from the expressions
given in Refs.~\cite{ERZ,B}, where all parameters should be
assumed to be given at the scale at which the matrix element
is evaluated, i.e. $M_{st}$ or $M_{sb}$.
As we have shown above,
the dominant D-term contributions to the Higgs masses are already
taken into account in the one-loop leading logarithmic
expressions, much as it happens in the case
of $m_Q = m_U$,  where
the dominant D-term contribution comes from the first two terms in
Eq.~(\ref{efflamix}). The D-term contribution of
$\Delta_{\rm th} \lambda_{\rm eff}$ becomes relevant only for
very large mixing and gives a very small correction to the
top quark Yukawa coupling effect. For completeness, we shall
include all ${\cal{O}}(g^2h_q^2,g'^2h_q^2)$ D-term contributions,
where $q = t,b$.

In order to obtain the expressions
for $\left(\widetilde{{\cal{M}}}^2_{ij}\right)_{\widetilde{t}} \;,
\left(\widetilde{{\cal{M}}}^2_{ij}\right)_{\widetilde{b}}$
one must subtract from
the expressions  of the matrix elements at the
scale $M_{st}$, $M_{sb}$, respectively,
the contributions coming from
the term dependent on the CP-odd Higgs mass and from the
leading order logarithmic expressions.
This procedure leaves in the matrix elements all the terms
that should be frozen at the scales $M_{st}$, $M_{sb}$, including
all terms depending on the squark mixing
mass parameters. Using the expressions given
in Refs.~\cite{ERZ,B}, and Eqs.~(\ref{lambimt})--(\ref{lambda4}),
we obtain,
\begin{eqnarray}
\label{cpevenst}
&&
\left(\widetilde{{\cal{M}}}_{ij}^2(M_{st})
\right)_{\widetilde{t}}
=
\frac{3}{8 \pi^2 v^2}
\left[
\begin{array}{cc}
\widetilde{\Delta}_{11}^{\widetilde{t}} +
\left(\widetilde{\Delta}'_{11}\right)^{\widetilde{t}}
& \widetilde{\Delta}_{12}^{\widetilde{t}} +
\left(\widetilde{\Delta}'_{12}\right)^{\widetilde{t}} \\
\widetilde{\Delta}_{12}^{\widetilde{t}} +
\left(\widetilde{\Delta}'_{12}\right)^{\widetilde{t}}
& \widetilde{\Delta}_{22}^{\widetilde{t}} +
\left(\widetilde{\Delta}'_{22}\right)^{\widetilde{t}}
\end{array}
\right],
\end{eqnarray}
\begin{eqnarray}
\label{cpevensb}
&&
\left(\widetilde{{\cal{M}}}_{ij}^2 (M_{sb})
\right)_{\widetilde{b}}
=
\frac{3}{8 \pi^2 v^2}
\left[
\begin{array}{cc}
\widetilde{\Delta}_{11}^{\widetilde{b}} +
\left(\widetilde{\Delta}'_{11}\right)^{\widetilde{b}}
& \widetilde{\Delta}_{12}^{\widetilde{b}} +
\left(\widetilde{\Delta}'_{12}\right)^{\widetilde{b}} \\
\widetilde{\Delta}_{12}^{\widetilde{b}} +
\left(\widetilde{\Delta}'_{12}\right)^{\widetilde{b}}
& \widetilde{\Delta}_{22}^{\widetilde{b}} +
\left(\widetilde{\Delta}'_{22}\right)^{\widetilde{b}}
\end{array}
\right],
\end{eqnarray}
where
\begin{eqnarray}
\label{delta11b}
\widetilde{\Delta}_{11}^{\widetilde{b}}& = &
\frac{m_b^4}{\cos^2\beta}
\left[\log \left(\frac{m_{\;\widetilde{b}_1}^2 m_{\;\widetilde{b}_2}^2}
{\left(m_Q^2 + m_b^2\right)\left(m_D^2 + m_b^2\right)}\right)
+\frac{2 A_b(A_b-\mu\tan\beta)}
{m_{\;\widetilde{b}_1}^2-m_{\;\widetilde{b}_2}^2}
\log\frac{m_{\;\widetilde{b}_1}^2}{m_{\;\widetilde{b}_2}^2}\right]
\nonumber \\
& + & \frac{m_b^4}{\cos^2\beta}
\left[\frac{A_b(A_b-\mu\tan\beta)}
{m_{\;\widetilde{b}_1}^2-m_{\;\widetilde{b}_2}^2}
\right]^2
g(m_{\;\widetilde{b}_1}^2,m_{\;\widetilde{b}_2}^2),
\end{eqnarray}
\begin{eqnarray}
\label{delta11t}
\widetilde{\Delta}_{11}^{\widetilde{t}}& = &
\frac{m_t^4}{\sin^2\beta}
\left[\frac{\mu(-A_t+\mu\cot\beta)}
{m_{\;\widetilde{t}_1}^2-m_{\;\widetilde{t}_2}^2}
\right]^2
g(m_{\;\widetilde{t}_1}^2,m_{\;\widetilde{t}_2}^2),
\end{eqnarray}
\begin{eqnarray}
\label{delta22t}
\widetilde{\Delta}_{22}^{\widetilde{t}} & = &
\frac{m_t^4}{\sin^2\beta}
\left[\log \left(\frac{m_{\;\widetilde{t}_1}^2 m_{\;\widetilde{t}_2}^2}
{\left(m_Q^2 + m_t^2\right)\left(m_U^2 + m_t^2\right)} \right)
+\frac{2 A_t(A_t-\mu\cot\beta)}
{m_{\;\widetilde{t}_1}^2-m_{\;\widetilde{t}_2}^2}
\log\frac{m_{\;\widetilde{t}_1}^2}
{m_{\;\widetilde{t}_2}^2}\right] \nonumber \\
& + & \frac{m_t^4}{\sin^2\beta}
\left[\frac{A_t(A_t-\mu\cot\beta)}{m_{\;\widetilde{t}_1}^2-
m_{\;\widetilde{t}_2}^2}
\right]^2
g(m_{\;\widetilde{t}_1}^2,m_{\;\widetilde{t}_2}^2),
\end{eqnarray}
\begin{eqnarray}
\label{delta22b}
\widetilde{\Delta}_{22}^{\widetilde{b}} & = &
\frac{m_b^4}{\cos^2\beta}
\left[\frac{\mu(-A_b+\mu\tan\beta)}{m_{\;\widetilde{b}_1}^2-
m_{\;\widetilde{b}_2}^2}
\right]^2
g(m_{\;\widetilde{b}_1}^2,m_{\;\widetilde{b}_2}^2),
\end{eqnarray}
\begin{eqnarray}
\label{delta12t}
\widetilde{\Delta}_{12}^{\widetilde{t}} & = &
\frac{m_t^4}{\sin^2\beta}
\frac{\mu(-A_t+\mu\cot\beta)}
{m_{\;\widetilde{t}_1}^2-m_{\;\widetilde{t}_2}^2}
\left[\log\frac{m_{\;\widetilde{t}_1}^2}{m_{\;\widetilde{t}_2}^2}
+\frac{A_t(A_t-\mu\cot\beta)}
{m_{\;\widetilde{t}_1}^2-m_{\;\widetilde{t}_2}^2}
g(m_{\;\widetilde{t}_1}^2,m_{\;\widetilde{t}_2}^2)\right],
\nonumber\\
\end{eqnarray}
\begin{eqnarray}
\label{delta12b}
\widetilde{\Delta}_{12}^{\widetilde{b}} & = &
\frac{m_b^4}{\cos^2\beta}
\frac{\mu(-A_b+\mu\tan\beta)}
{m_{\;\widetilde{b}_1}^2-m_{\;\widetilde{b}_2}^2}
\left[\log\frac{m_{\;\widetilde{b}_1}^2}{m_{\;\widetilde{b}_2}^2}
+\frac{A_b(A_b-\mu\tan\beta)}
{m_{\;\widetilde{b}_1}^2-m_{\;\widetilde{b}_2}^2}
g(m_{\;\widetilde{b}_1}^2,m_{\;\widetilde{b}_2}^2)\right],
\nonumber\\
\end{eqnarray}
and
\begin{eqnarray}
\left(\widetilde{\Delta}'_{11}\right)^{\widetilde{b}}
& = & M_Z^2 \left[ 2 m_b^2
f^{\;\widetilde{b}}_1
- m_b A_b f^{\;\widetilde{b}}_2 \right], \;\;\;\;\;\;\;\;\;\;
\left(\widetilde{\Delta}'_{11}\right)^{\widetilde{t}}  =  M_Z^2
m_t \mu \cot\beta f^{\;\widetilde{t}}_2\;
\nonumber\\
\left(\widetilde{\Delta}'_{22}\right)^{\widetilde{t}}
& = & M_Z^2 \left[ -2 m_t^2
f^{\;\widetilde{t}}_1
+ m_t A_t f^{\;\tilde{t}}_2 \right], \;\;\;\;\;\;\;\;\;
\left(\widetilde{\Delta}'_{22}\right)^{\widetilde{b}}  =  - M_Z^2
m_b \mu \tan\beta f^{\;\widetilde{b}}_2\;
\nonumber\\
\left(\widetilde{\Delta}'_{12}\right)^{\widetilde{t}}
& = & M_Z^2 \left[ m_t^2 \cot\beta
f^{\;\widetilde{t}}_1
-  m_t \frac{ A_t \cot\beta + \mu}{2}
f^{\;\widetilde{t}}_2 \right],
\nonumber\\
\left(\widetilde{\Delta}'_{12}\right)^{\widetilde{b}} & = & M_Z^2 \left[
- m_b^2 \tan\beta
f^{\;\widetilde{b}}_1
+  m_b \frac{A_b \tan\beta + \mu}{2} f^{\;\widetilde{b}}_2 \right],
\end{eqnarray}
where
\begin{equation}
g(m_1^2,m_2^2)=2-\frac{m_1^2+m_2^2}{m_1^2-m_2^2}
\log\frac{m_1^2}{m_2^2},
\end{equation}

\begin{eqnarray}
f^{\;\widetilde{t}}_1 & = & \frac{m_Q^2 - m_U^2}
{m_{\;\widetilde{t}_1}^2 - m_{\;\widetilde{t}_2}^2}
\left(\frac{1}{2} - \frac{4}{3}
\sin^2\theta_W\right) \log \left(
\frac{m_{\;\widetilde{t}_1}}{m_{\;\widetilde{t}_2}}\right)
\nonumber\\
&+& \left(\frac{1}{2} - \frac{2}{3} \sin^2\theta_W \right)
\log\left(\frac{m_{\;\widetilde{t}_1} m_{\;\widetilde{t}_2}}{m_Q^2 + m_t^2}
\right)
\\
& + & \frac{2}{3} \sin^2\theta_W \log \left(
\frac{m_{\;\widetilde{t}_1}m_{\;\widetilde{t}_2}}
{m_U^2 + m_t^2} \right) \nonumber,
\end{eqnarray}
\begin{eqnarray}
f^{\;\widetilde{b}}_1 & = & \frac{m_Q^2 - m_D^2}
{m_{\;\widetilde{b}_1}^2 - m_{\;\widetilde{b}_2}^2}
\left(-\frac{1}{2} + \frac{2}{3}
\sin^2\theta_W\right) \log \left(
\frac{m_{\;\widetilde{b}_1}}{m_{\;\widetilde{b}_2}}\right)
\nonumber\\
&+& \left(-\frac{1}{2} + \frac{1}{3} \sin^2\theta_W \right)
\log\left(\frac{m_{\;\widetilde{b}_1} m_{\;\widetilde{b}_2}}{m_Q^2 + m_b^2}
\right)
\\
&-& \frac{1}{3} \sin^2\theta_W \log \left(
\frac{m_{\;\widetilde{b}_1}m_{\;\widetilde{b}_2}}
{m_D^2 + m_b^2} \right)\nonumber,
\end{eqnarray}
\begin{eqnarray}
f^{\;\widetilde{t}}_2 & = &
m_t\; \frac{A_t - \mu \cot\beta}{m_{\;\widetilde{t}_1}^2 -
m_{\;\widetilde{t}_2}^2} \left[-\frac{1}{2}
\log\left( \frac{m_{\;\widetilde{t}_1}^2}
{m_{\;\widetilde{t}_2}^2}\right)\right. \nonumber \\
& + &\left.
\left(\frac{4}{3} \sin^2\theta_W
- \frac{1}{2} \right)
\frac{m_Q^2 - m_U^2}
{m_{\;\widetilde{t}_1}^2 -  m_{\;\widetilde{t}_2}^2}
g(m_{\;\widetilde{t}_1}^2, m_{\;\widetilde{t}_2}^2) \right],
\end{eqnarray}

\begin{eqnarray}
f^{\;\widetilde{b}}_2 & = &
m_b\; \frac{A_b - \mu \tan\beta}{m_{\;\widetilde{b}_1}^2 -
m_{\;\widetilde{b}_2}^2}
 \left[ \frac{1}{2} \log\left( \frac{m_{\;\widetilde{b}_1}^2}
{m_{\;\widetilde{b}_2}^2}\right) \right. \nonumber \\
& +  & \left.
\left( \frac{1}{2} -\frac{2}{3} \sin^2\theta_W
 \right)
\frac{m_Q^2 - m_D^2}
{m_{\;\widetilde{b}_1}^2 -  m_{\;\widetilde{b}_2}^2}
g(m_{\;\widetilde{b}_1}^2, m_{\;\widetilde{b}_2}^2) \right].
\end{eqnarray}
In the above, all terms should be computed at the scale
$M_{st}$ or $M_{sb}$ depending on whether they are associated to
stop or sbottom contributions
and all ${\cal{O}}(g^4,g^2g'^2,g'^4)$ terms are ignored.
It is easy to show that, apart from very small terms of
order $M_Z^2/(m_{\;\widetilde{t}_1}^2 + m_{\;\widetilde{t}_2}^2)$,
Eqs.~(\ref{cpevenst}), (\ref{cpevensb})
vanish  in the case of zero squark mixing.
It is also straightforward to show that in the limit of large
$\tan\beta$, $m_Q = m_U$  and $m_b = \mu =0$,
$\widetilde{\Delta}_{12}^{\widetilde{t}} = 0$
and $3 \left(\widetilde{\Delta}_{22}^{\widetilde{t}} +
\left(\widetilde{\Delta}_{22}'\right)^{\widetilde{t}}
\right)/16\pi^2v^4$
reproduces the expression of $\Delta_{\rm th} \lambda_{\rm eff}$
given in Eq.~(\ref{umbral}).
All parameter values in the expressions
for the matrix elements $\widetilde{{\cal{M}}}_{ij}^2$
can be expressed in terms of
their values at the scale $\mbart$ by making use of the corresponding
$\beta$-and $\gamma$-functions.

Having computed the renormalization group improved Higgs mass
matrix elements at the scale $\mbart$, the neutral CP-even Higgs
mass eigenvalues can be easily derived. They read
 \begin{eqnarray}
m^2_{h(H)} &=& \frac{{\rm Tr} {\cal M}^2 \mp \sqrt{({\rm Tr}
{\cal M}^2)^2 - 4 \det {\cal M}^2}}{2}
\label{mhH}
\end{eqnarray}
where
\begin{equation}
   {\rm Tr}{\cal M}^2 = {\cal{M}}_{11}^2 + {\cal{M}}_{22}^2 \;\; ; \;\;\;\;\;
     \det {\cal M}^2 = {\cal{M}}_{11}^2 {\cal{M}}_{22}^2 -
     \left( {\cal{M}}_{12}^2 \right)^2 ,
\label{detm2}
\end{equation}
and the ${\cal{M}}_{ij}$ are the renormalized Higgs mass matrix elements
at the scale $\mbart$, Eq.~(\ref{m2ijmt}).
{}From the matrix elements,
the mixing angle $\alpha$ is also determined by~\cite{HH}:
\begin{eqnarray}
      \sin 2\alpha = \frac{2{\cal{M}}_{12}^2}
{\sqrt{\left({\rm Tr} {\cal M}^2\right)^2-4\det {\cal M}^2}}
 \end{eqnarray}
\begin{eqnarray}
      \cos 2\alpha = \frac{{\cal{M}}_{11}^2-{\cal{M}}_{22}^2}
{\sqrt{\left({\rm Tr} {\cal M}^2\right)^2-4\det {\cal M}^2}}
\end{eqnarray}

Concerning the running of the squark mass parameters, the
dominant contribution comes from gluino-induced effects, which
are absent in the case of heavy gluino particles, as we are
considering within this work. The remaining contributions
are small and have a somewhat complicated dependence on the squark and
Higgs spectrum. We shall ignore them within
our approximation. We have further defined the light stop and sbottom
masses as the values obtained using Eqs. (\ref{masast}), (\ref{masasb}),
while taking the running mass parameters at the scale $\mbart$
and adding the QCD-dependent vacuum polarization effects. A more
precise definition of the squark masses may  be obtained
by computing the squark effective potential and adding the
full vacuum polarization contributions to the squark masses,
much as we have done in the case of the Higgs bosons. We shall
concentrate on this subject elsewhere.

Figures~3 and 4 show the dependence of the Higgs mass $m_h$
for varying values of $m_Q = A_t = m_A$ and
for a fixed
value of the right-handed mass parameters $m_U = m_D =1$ TeV and
$m_U = m_D = 100$ GeV, respectively. The supersymmetric
mass parameter $\mu$ has been set to zero. The solid lines
represent the value of the Higgs mass by performing the
renormalization group improvement of the effective potential
method, as explained in this work. The dotted lines represent
the values obtained from the
one-loop effective potential, while ignoring the stop decoupling
and taking all values to be given at the scale $\mbart$
\cite{ERZ}. The dashed
lines are the values obtained for the Higgs mass, while considering
that the effective potential is scale-invariant, that is by
taking the second derivatives at the scale $M_S$ and rescaling
them to the scale $\mbart$ through the appropriate anomalous dimension
factors (see e.g. Ref.~\cite{CEQW}).
For low values of $m_U$ and $m_D$, as those shown in
Fig. 4, the last method becomes accurate for all values of
$m_Q < 600$ GeV, while in the second method the departure from the
proper renormalization group improved values is faster. Observe that
the behaviour shown in Fig. 4 is very similar to the one shown
in Fig. 1. Since the main purpose of Figs.~3 and 4 is to compare
the results in this paper with
other different approaches,
based on the one-loop MSSM effective potential, we have
plotted in them the running Higgs mass at the scale $\overline{m}_t$,
and turned the small D-terms threshold corrections off,
i.e. we have put $\widetilde{\Delta}'_{ij}=0$.
In the following we shall turn these D-terms on and consider
pole Higgs masses.

The Higgs masses $m_{h,H,A}$ defined in Eq.~(\ref{mhH}) are all running
masses obtained from the effective potential\footnote{In fact, as was
noticed in Ref.~\cite{ERZ}, the mass of the CP-odd Higgs, $m_A$,
turns out to be scale independent at one loop.}, and evaluated at
the top-quark mass scale. To compute the physical (propagator
pole) masses $M_{h,H,A}$ one has to correct for the fact that the
effective potential is defined at zero external momentum. In fact, the
pole and running Higgs masses are related by (see e.g. Ref.~\cite{CEQR})
\be
\label{polemasas}
M^2_{\exis}=m^2_{\exis}+{\rm Re}\Delta\Pi_{\exis}(M^2_{\exis})
\ee
where ${\exis}=h,H,A$ and
\be
\label{delpolariz}
\Delta\Pi_{\exis}(M^2_{\exis})=\Pi(M^2_{\exis})-\Pi(0) ,
\ee
$\Pi_{\exis}(q^2)$ being the renormalized self-energy
of the corresponding Higgs boson.
We have computed the Higgs self-energies, at the one-loop
level, from the top/stop and bottom/sbottom sectors.  The
corresponding expressions can be found in Appendix A.

Figures~5--7 show the variation of the pole Higgs mass $M_h$
as a function of $m_Q$ for
fixed $m_A$ and several fixed values of $m_U$ and of the stop
mixing parameter $A_t$. Although in general, for
a fixed moderate value of $m_U$ and a fixed value of $A_t$, the
Higgs mass increases together with $m_Q$,
for large values of $A_t$ and
moderate values of $m_U$  or
for small values of
$m_U$ and moderate values of $A_t$,
situations can be observed for which the Higgs mass
decreases with larger values of $m_Q$.
An understanding
of this effect may be obtained by making use of the approximation
of Ref.~\cite{CEQW}: although
$\log(M_{\rm SUSY}/\mbart)$, with
$M_{\rm SUSY}^2 \equiv
 \left(m_{\;\widetilde{t}_1}^2 + m_{\;\widetilde{t}_2}^2 \right)/2$,
increases together with the
squark-mass parameters, leading to larger values of the
Higgs mass, the squark mixing contributions are
maximized for values of $A_t^{\rm max} \simeq 2.4 M_{\rm SUSY}$. Hence,
for fixed values of $A_t$ a delicate balance between these
two effects should be present in order to maximize the Higgs
mass.  Observe also that for small values of $m_U$, such as
the ones analysed in Fig.~7, the configurations which
maximize the lightest CP-even Higgs mass  $M_h$
correspond to situations
for which one of the stops may become light. Indeed, the
curves are cut since we have introduced the (crude) experimental
constraint $m_{\st_{2}}\simgt 45$ GeV on them.
An important effect arising in the case of small values
of one of the mass parameters, for instance $m_U^2 \simeq M_Z^2$,
and $m_Q^2 \gg m_U^2$ as shown in Fig.~7, is that for the same
value of $M_{\rm SUSY}$, the maximal value for the Higgs mass is always
lower than in the case of $m_Q = m_U$. This is due to the fact that
the requirement of having stop masses above the present
experimental bound implies $A_t \leq m_Q \simeq \sqrt{2} M_{\rm SUSY}$.
Hence, $A_t$ is always significantly lower than  $A_t^{\rm max}$,
which, as explained above, is the value that maximizes the Higgs mass
for that particular value of the average stop mass scale $M_{\rm SUSY}$.

One could enquire about the stop and sbottom
vacuum polarization contributions in the pole Higgs boson mass definitions.
We have checked that these contributions do not give a significant
effect in the determination of the neutral Higgs boson masses, unless one
of the squarks becomes light (i.e. $\st$ and/or $\sb$) and its couplings
to the Higgs fields
(i.e. $A_t$, $A_b$, $\mu$) are large. This behaviour is illustrated in
Fig.~8 where we plot the neutral Higgs boson running
($m_h$, $m_H$, $m_A$) masses (dotted lines) and pole ($M_h$, $M_H$,
$M_A$) masses (solid lines) as functions of $A_t$, for fixed values of the
other supersymmetric parameters. We also plot the mass of the lightest
stop $m_{\st_{2}}$ (dashed line).
We see that the departure between the running
masses and the pole masses occurs in all cases
for large values of $A_t$ and
small values of $m_{\st_{2}}$.

The dependence of the Higgs mass on $m_D$ becomes relevant only
for large values of $\tan\beta$.
In Fig.~9 we show the variation of $M_h$
as a function of $m_D$ for $\tan\beta=60$, fixed values of $m_Q$, $m_U$
and $m_A$, and two values of $\mu$: $\mu=1$ TeV (solid curves) and
$\mu=2$ TeV (dashed curves). The radiative corrections
to $M_h$ induced by the
bottom/sbottom propagation are always of negative sign and become
only relevant for very large values of the $\mu$-parameter.
The behaviour with $\mu$ can be understood
from the fact that the larger $\mu$, the larger mixing in the sbottom sector
and hence lighter sbottom masses are obtained.
In this example, we have chosen $m_b(\mbart) = 3$ GeV, which corresponds
to a bottom quark pole mass $M_b \simeq 5$ GeV.
All the curves are cut by the experimental
constraint $m_{\sb_{2}}\simgt 45$ GeV.
Finally Fig.~9 also exhibits the dependence of the Higgs mass on the
parameter $A_t$, its dependence with respect to $A_b$ being tiny.

\nsect{Conclusions}

We have presented a renormalization group improvement of the
effective potential computation of the neutral Higgs masses in the
MSSM. The method provides the first calculation of two-loop
leading order corrections to the Higgs masses valid for
any value of the soft supersymmetry breaking squark mass parameters,
$m_Q$, $m_U$, $m_D$, $A_t$ and $A_b$, the CP-odd mass $m_A$,
the supersymmetric Higgs mass $\mu$
and $\tan\beta$.  This generalization is essential for the
computation of the Higgs masses and mixing angles in the presence
of light squarks. Our method uses
explicit decoupling of stops and sbottoms
at their corresponding mass scales, leaving threshold effects in the
effective potential (and coupling constants)
frozen at the decoupling scales and evolving, in the
squared mass matrix, with the anomalous dimensions of the Higgs
fields.

The threshold effects achieve a complete matching of the
effective potential for scales above and below  the decoupling
scales, and include all higher order (non-renormalizable)
terms arising from the whole MSSM effective
potential. The effect of considering non-renormalizable threshold
effects in the effective potential is twofold: on the one hand
it triggers resummations in
the renormalization group expansion of the parameters,
leading to `physical' expansion parameters; on the other hand,
it enables to consider the general case of
arbitrary left--right squark mixing,
as well as general left- and right-handed soft supersymmetry breaking
squark masses.

We have corrected the running neutral Higgs boson masses with
one-loop self-energy diagrams, where top- and bottom-quarks,
stops and sbottoms propagate, to define the corresponding
pole masses. The numerical effect of polarizations is relevant only
under special circumstances: light squarks and large mixing.

We have analysed
the general pattern of Higgs
masses for general values of the supersymmetric parameters.
We have found regions in the parameter space where the
radiative corrections become large and negative. They are
characterized by large values of the mixing-mass parameters,
where the stability of the electroweak minimum can be endangered
by the presence of charge and color breaking minima.
Our results also allow the evaluation of the relevant
radiatively corrected Higgs couplings through the corresponding
value of the Higgs angle $\alpha$ \cite{CEQW}.

We have neglected, throughout the whole calculation, the possible
contribution coming from light charginos/neutralinos. Their effect
can be easily included in the threshold terms, as well as in the
running of the $\beta$- and $\gamma$-functions, where they appear
as ${\cal O}(g^4,g^2g'^2,g'^4)$
terms and are thus numerically unimportant.
In Ref.~\cite{CEQW} we have shown that, in the case of a heavy
supersymmetric spectrum, our analytical expressions reproduced
the Higgs mass spectrum with an
error of less than 2--3 GeV.
It can be easily  checked that  light charginos and neutralinos
can increase the Higgs
mass in $\simlt$ 2--3 GeV \cite{topc,Marc}, with respect to heavy ones,
which is indeed
within the errors of our different approximations.
Nevertheless for completeness, we include in appendix B
the leading-log ${\cal{O}}(g^4,g'^2g^2,g'^4)$
chargino and  neutralino  contributions to the CP-even Higgs
masses and the mixing angle. Throughout this work,
we have also implicitly assumed that the gluino masses are of
order $M_S$. If the gluinos were, instead, much lighter than the
characteristic squark masses, the running of the
third generation Yukawa couplings
would be different, inducing a small indirect effect
on the two-loop Higgs mass computation. The third generation
Yukawa coupling running is also modified by the presence of
light charginos/neutralinos in the spectrum. These two loop
contributions to the CP-even Higgs masses are also presented
in appendix B.

Our present analysis reproduces, with a high level
of accuracy, the values of
the Higgs masses and mixing angles, for the
previously studied case of
degenerate left- and right-handed squark mass parameters,
and for values of the squark left--right mixing
mass parameters lower than the ones giving the maximal
values of the lightest CP-even Higgs mass. This comparison
holds up to a tiny difference coming from the inclusion in this
work of the small D-term threshold contributions and
vacuum-polarization effects. This confirms
previous results on the upper bound on the lightest Higgs boson
mass in the MSSM.

\section*{Acknowledgements}

We would like to thank
A. Brignole, J.R. Espinosa, H. Haber, S. Peris
and F. Zwirner for interesting discussions.

\newpage

\appendixA{Appendix A}

In this appendix we  provide the analytical expressions for the
Higgs-boson self-energies in Eq.~(\ref{delpolariz}).
We first define
the different self-energy contributions as
\be
\label{sumapol}
\Delta\Pi_{\exis}(M_{\exis}^2)=\Delta\Pi_{\exis}^{(t)}(M_{\exis}^2)
+\Delta\Pi_{\exis}^{(b)}(M_{\exis}^2)
+\Delta\Pi_{\exis}^{(\st\;)}(M_{\exis}^2)
+\Delta\Pi_{\exis}^{(\sb\;)}(M_{\exis}^2)
\ee
where ${\exis}=h,H,A$. The different contributions in
(A.1) read:
\be
\label{ht}
\Delta\Pi^{(t)}_h(M_h^2)=\frac{3}{8\pi^2}h_t^2\cos^2\alpha
\left[-2M_t^2+\frac{1}{2}M_h^2\right]f(M_t^2,M_t^2,M_h^2)
\ee
\be
\label{Ht}
\Delta\Pi^{(t)}_H(M_H^2)=\frac{3}{8\pi^2}h_t^2\sin^2\alpha
\left[-2M_t^2+\frac{1}{2}M_H^2\right]f(M_t^2,M_t^2,M_H^2)
\ee
\be
\label{At}
\Delta\Pi^{(t)}_A(M_A^2)=\frac{3}{8\pi^2}h_t^2\cos^2\beta
\left[-\frac{1}{2}M_A^2\right]f(M_t^2,M_t^2,M_A^2)
\ee
\be
\label{hb}
\Delta\Pi^{(b)}_h(M_h^2)=\frac{3}{8\pi^2}h_b^2\sin^2\alpha
\left[-2m_b^2+\frac{1}{2}M_h^2\right]f(m_b^2,m_b^2,M_h^2)
\ee
\be
\label{Hb}
\Delta\Pi^{(b)}_H(M_H^2)=\frac{3}{8\pi^2}h_b^2\cos^2\alpha
\left[-2m_b^2+\frac{1}{2}M_H^2\right]f(m_b^2,m_b^2,M_H^2)
\ee
\be
\label{Ab}
\Delta\Pi^{(b)}_A(M_A^2)=\frac{3}{8\pi^2}h_b^2\sin^2\beta
\left[-\frac{1}{2}M_A^2\right]f(m_b^2,m_b^2,M_A^2)
\ee
\be
\label{xstop}
\Delta\Pi_{\exis}^{(\st\;)}(M_{\exis}^2)=\sum_{i,j=1}^2
\frac{3}{16\pi^2}\left|C_{{\varphi}ij}^{(\st\;)}\right|^2
f(\msti^2,\mstj^2,M_{\exis}^2)
\ee
\be
\label{xsbottom}
\Delta\Pi_{\exis}^{(\sb\;)}(M_{\exis}^2)=\sum_{i,j=1}^2
\frac{3}{16\pi^2}\left|C_{{\varphi}ij}^{(\sb\;)}\right|^2
f(\msbi^2,\msbj^2,M_{\exis}^2)
\ee
The different coefficients in (A.8) and (A.9)
are:
\bea
\label{hstop}
C_{hij}^{(\st\;)} & = &
\frac{2 \sqrt{2} \sin^2\theta_W}{3} \frac{M_Z^2}{v}
\sin(\beta+\alpha) \left[ \delta_{ij} + \frac{ 3 - 8
\sin^2\theta_W}{4 \sin^2\theta_W} Z_U^{1i} Z_U^{1j} \right]
\\
& - & \sqrt{2} h_t^2v\sin\beta\cos\alpha \ \delta_{ij}
-\frac{1}{\sqrt{2}}h_t(A_t\cos\alpha+\mu\sin\alpha)
(Z_U^{1i *}Z_U^{2j}+Z_U^{1j}Z_U^{2i *}) \nonumber
\eea
\bea
\label{Hstop}
C_{Hij}^{(\st\;)}&=&
-\frac{2 \sqrt{2} \sin^2\theta_W}{3} \frac{M_Z^2}{v}
\cos(\beta+\alpha) \left[ \delta_{ij} + \frac{ 3 - 8
\sin^2\theta_W}{4 \sin^2\theta_W} Z_U^{1i} Z_U^{1j} \right]
\\
&-&\sqrt{2}h_t^2 v\sin\beta\sin\alpha \ \delta_{ij}
-\frac{1}{\sqrt{2}}h_t(A_t\sin\alpha-\mu\cos\alpha)
(Z_U^{1i *}Z_U^{2j}+Z_U^{1j}Z_U^{2i *})\nonumber
\eea
\be
\label{Astop}
C_{Aij}^{(\st\;)}=
-\frac{1}{\sqrt{2}}h_t(A_t\cos\beta+\mu\sin\beta)
(Z_U^{1i *}Z_U^{2j}-Z_U^{1j}Z_U^{2i *})
\ee
\bea
\label{hsbottom}
C_{hij}^{(\sb\;)}&=&
-\frac{ \sqrt{2} \sin^2\theta_W}{3} \frac{M_Z^2}{v}
\sin(\beta+\alpha) \left[ \delta_{ij} + \frac{ 3 - 4
\sin^2\theta_W}{2 \sin^2\theta_W} Z_D^{1i} Z_D^{1j} \right]
\\
& + & \sqrt{2}h_b^2v\cos\beta\sin\alpha \ \delta_{ij}
+\frac{1}{\sqrt{2}}h_b(A_b\sin\alpha+\mu\cos\alpha)
(Z_D^{1i *}Z_D^{2j}+Z_D^{1j}Z_D^{2i *})\nonumber
\eea
\bea
\label{Hsbottom}
C_{Hij}^{(\sb\;)}&=&
\frac{ \sqrt{2} \sin^2\theta_W}{3} \frac{M_Z^2}{v}
\cos(\beta+\alpha) \left[ \delta_{ij} + \frac{ 3 - 4
\sin^2\theta_W}{2 \sin^2\theta_W} Z_D^{1i} Z_D^{1j} \right]
\\
& - & \sqrt{2}h_b^2v\cos\beta\cos\alpha \ \delta_{ij}
-\frac{1}{\sqrt{2}}h_b(A_b\cos\alpha-\mu\sin\alpha)
(Z_D^{1i *}Z_D^{2j}+Z_D^{1j}Z_D^{2i *}) \nonumber
\eea
\be
\label{Asbottom}
C_{Aij}^{(\sb\;)}=
\frac{1}{\sqrt{2}}h_b(A_b\sin\beta+\mu\cos\beta)
(Z_D^{1i *}Z_D^{2j}-Z_D^{1j}Z_D^{2i *}),
\ee
where the matrices $Z_U^{ij}$ and $Z_D^{ij}$ are those diagonalizing the
stop and sbottom squared mass matrices,
(\ref{masastop}) and (\ref{masasbottom}),
respectively
\be
\label{diagt}
Z^{\dagger}_U M^2_{\st}Z_U=
\left(
\begin{array}{cc}
m_{\; \widetilde{t}_1} ^2 & 0 \\
0 & m_{\; \widetilde{t}_2} ^2
\end{array}
\right)
\ee
\be
\label{diagb}
Z^{\dagger}_D M^2_{\sb}Z_D=
\left(
\begin{array}{cc}
m_{\; \widetilde{b}_1} ^2 & 0 \\
0 & m_{\; \widetilde{b}_2} ^2
\end{array}
\right).
\ee

The function
\be
\label{integrall}
f(m_1^2,m_2^2,q^2)=\int_0^1 dx \log\frac{m_1^2(1-x)+m_2^2 x-q^2 x(1-x)}
{m_1^2(1-x)+m_2^2 x},
\ee
which arises from the integration of the
loop of (scalar) particles, is given by
\be
\label{funcion}
f(m_1^2,m_2^2,q^2)=-1+\frac{1}{2}
\left(\frac{m_1^2+m_2^2}{m_1^2-m_2^2}-\ \delta\right)
\log\frac{m_2^2}{m_1^2}+\frac{1}{2}r\log\left[\frac{(1+r)^2-\ \delta^2}
{(1-r)^2-\ \delta^2}\right]
\ee
with
\be
\label{deltaa}
\ \delta=\frac{m_1^2-m_2^2}{q^2}
\ee
and
\be
\label{ere}
r=\sqrt{(1+\ \delta)^2-\frac{4m_1^2}{q^2}}
\ee

\appendixB{Appendix B}

In this appendix we present the leading-logarithmic
${\cal{O}}(g^4,g^2g'^2,g'^4)$  chargino
and neutralino contributions to the CP-even Higgs masses and
mixing angle. We first present the
modification  of the quartic couplings, Eq.~(\ref{lambimt}),
obtained while changing the chargino
and neutralino masses from $M_{\rm SUSY}$
to an overall mass $m_{\;\tilde{\chi}}$. We get,
\begin{eqnarray}
\Delta \lambda_1 = \Delta \lambda_2 & = &
\left(\frac{9}{64 \pi^2} g^4 + \frac{5}{192 \pi^2} g'^4
\right)  \log\left( \frac{M_{\rm SUSY}^2}{m_{\;\tilde{\chi}}^2} \right)
\nonumber \\
\label{delambi}
\Delta (\lambda_3 + \lambda_4) &=
 & \left( \frac{3}{ 64 \pi^2} g^4 + \frac{7}{192 \pi^2} g'^4
+ \frac{1}{8 \pi^2} g'^2 g^2 \right)
\log\left( \frac{M_{\rm SUSY}^2}{m_{\;\tilde{\chi}}^2} \right),
\end{eqnarray}
where in the case of charginos lighter that the top quark mass
one has to replace $m_{\;\tilde{\chi}}^2$ by $\overline{m}_t^2$.
{}From Eq.~(B.1) the corresponding contributions to
the Higgs mass matrix  elements can be obtained:
\begin{eqnarray}
 \Delta  {\cal{M}}_{11}^2 & = & 2 \Delta \lambda_1 v^2 \cos^2 \beta,
\;\;\;\;\;\;
\Delta  {\cal{M}}_{22}^2  = 2 \Delta \lambda_2 v^2 \sin^2 \beta
\nonumber\\
\Delta  {\cal{M}}_{12}^2   & = & \Delta{\cal{M}}_{21}^2 =
2 \Delta (\lambda_3 + \lambda_4) v^2 \sin \beta \cos \beta.
\label{b2}
\end{eqnarray}

Since the chargino and neutralino loops lead to small corrections
to the Higgs spectrum, it is possible to obtain  approximate
expressions for the CP-even Higgs masses, while keeping only
terms linear in the
mass matrix corrections $\Delta {\cal{M}}_{ij}^2$.
The corrections to the Higgs mass eigenvalues are given by
\begin{eqnarray}
\Delta m_{h,H}^2 = v^2 \left[\Delta \lambda_1 \mp  \left( \cos 2\alpha
\cos 2\beta \Delta \lambda_1 + \sin 2\alpha
\sin 2\beta \Delta (\lambda_3 + \lambda_4) \right) \right].
\label{delmh}
\end{eqnarray}
Analogously, the correction to the Higgs mixing
angle $\alpha$  reads
\begin{equation}
\Delta \alpha = \frac{v^2}{m_H^2 - m_h^2}
\left( - \sin 2\alpha
\cos 2\beta \Delta \lambda_1 + \cos 2\alpha
\sin 2\beta \Delta (\lambda_3 + \lambda_4) \right).
\label{delalpha}
\end{equation}
In the above, we have always assumed that the corrections to the
Higgs mass matrix elements are small in comparison not only to
the squared Higgs mass eigenvalues but also to their difference,
$m_H^2 - m_h^2$. If these
conditions are not fulfilled, the above approximate expressions,
Eqs.~(B.3) and (B.4) are no longer valid.
This might happen, for example, in the large
$\tan\beta$ regime, where for some particular value of the CP-odd mass,
the CP-even Higgs bosons become almost degenerate. In this case
the full
diagonalization of the Higgs mass matrix is needed in order to
get the correct expression for the Higgs mass eigenvalues and
mixing angle.

The presence of light gauginos and Higgsinos can also affect the
two loop corrections to the Higgs masses. This is due to the fact
that they affect the top and bottom Yukawa coupling beta functions,
which enter into the masses through the derivative of the
quartic coupling beta
functions, $\beta'_i$, Eq. (\ref{lambimt}), and through the
soft supersymmetry breaking and
fermion mass dependence in Eq. (\ref{m2ijmt}).
The most relevant effects
come from the possible presence of a gluino lighter than the
characteristic stop and sbottom masses, $m_{\widetilde{g}} < M_S$.
The dominant corrections to
$\beta'_i$ are given by
\begin{eqnarray}
\Delta \left(\frac{\beta'^{XY}_1}{2}\right)
& = &
\Delta \left( \frac{3}{32 \pi^2} \frac{{\rm d} h_b^4}
{{\rm d} \log Q^2} \right)
\left( 2 - \theta_Q
- \theta_D \right) \; \theta_t
\nonumber\\
&= &\frac{3}{\left(16\pi^2\right)^2} h_b^4 \left\{
\theta_{\widetilde{\chi}}
\left[ \frac{h_b^2}{2}
\left( 2 \theta_{Q} + \theta_{D} \right)
+ \frac{h_t^2}{2} \theta_{U} \right] \right.
\nonumber\\
& - & \left. \frac{4}{3} \left( \theta_{Q} + \theta_D \right)
g_3^2 \theta_{\widetilde{g}} \right\} \left( 2 - \theta_Q
- \theta_D \right) \; \theta_t
\end{eqnarray}
and
\begin{eqnarray}
\Delta \left(\frac{\beta'^{XY}_2}{2}\right) & = &
\Delta \left( \frac{3}{32 \pi^2} \frac{{\rm d} h_t^4}
{{\rm d}\log Q^2} \right)
\left( 2 - \theta_Q
- \theta_U \right) \; \theta_t
\nonumber\\
& = & \frac{3}{\left(16\pi^2\right)^2} h_t^4 \left\{
\theta_{\widetilde{\chi}} \left[ \frac{h_t^2}{2}
\left( 2 \theta_{Q} + \theta_{U} \right)
+ \frac{h_b^2}{2} \theta_{D} \right] \right.
\nonumber\\
& - & \left. \frac{4}{3} \left( \theta_{Q} + \theta_U \right)
g_3^2 \theta_{\widetilde{g}} \right\}
\left( 2 - \theta_Q - \theta_U \right) \; \theta_{t},
\end{eqnarray}
where $X,Y = Q,U,D,\widetilde{g},\widetilde{\chi},t$,
the superscript $XY$ indicating that we are
considering the behaviour of the quartic coupling between the
energy scales $m_X$ and $m_Y$ ($m_X > m_Y$) and the
value of the functions $\theta_i$
should be set to one if $m_i < m_X$ and
$\theta_i = 0$ otherwise.  The $\theta_{t}$ factor is included
to make explicit the fact that we are evaluating all quartic
couplings at the scale $\mbart$.
The two-loop corrections to the Higgs mass matrix elements
coming from the variation of the quartic couplings at $\mbart$
for the case of light gauginos and/or Higgsinos may
be obtained through Eq. (\ref{b2}), where
the variation of the quartic couplings may be obtained by
 a simple generalization of Eq.
(\ref{lambimt}) to the case under consideration:
\begin{equation}
\Delta \lambda_i = \sum_{XY} \Delta \left( \frac{\beta'^{XY}_i}{2}
\right)  \left( \widetilde{t}_X^{\;2} - \widetilde{t}_Y^{\;2} \right),
\end{equation}
where $XY$ denote pair of masses in hierarchical order and
$\widetilde{t}_i = \log(m_i^2/\mbart^2)$.

\begin{figure}
\centerline{
\psfig{figure=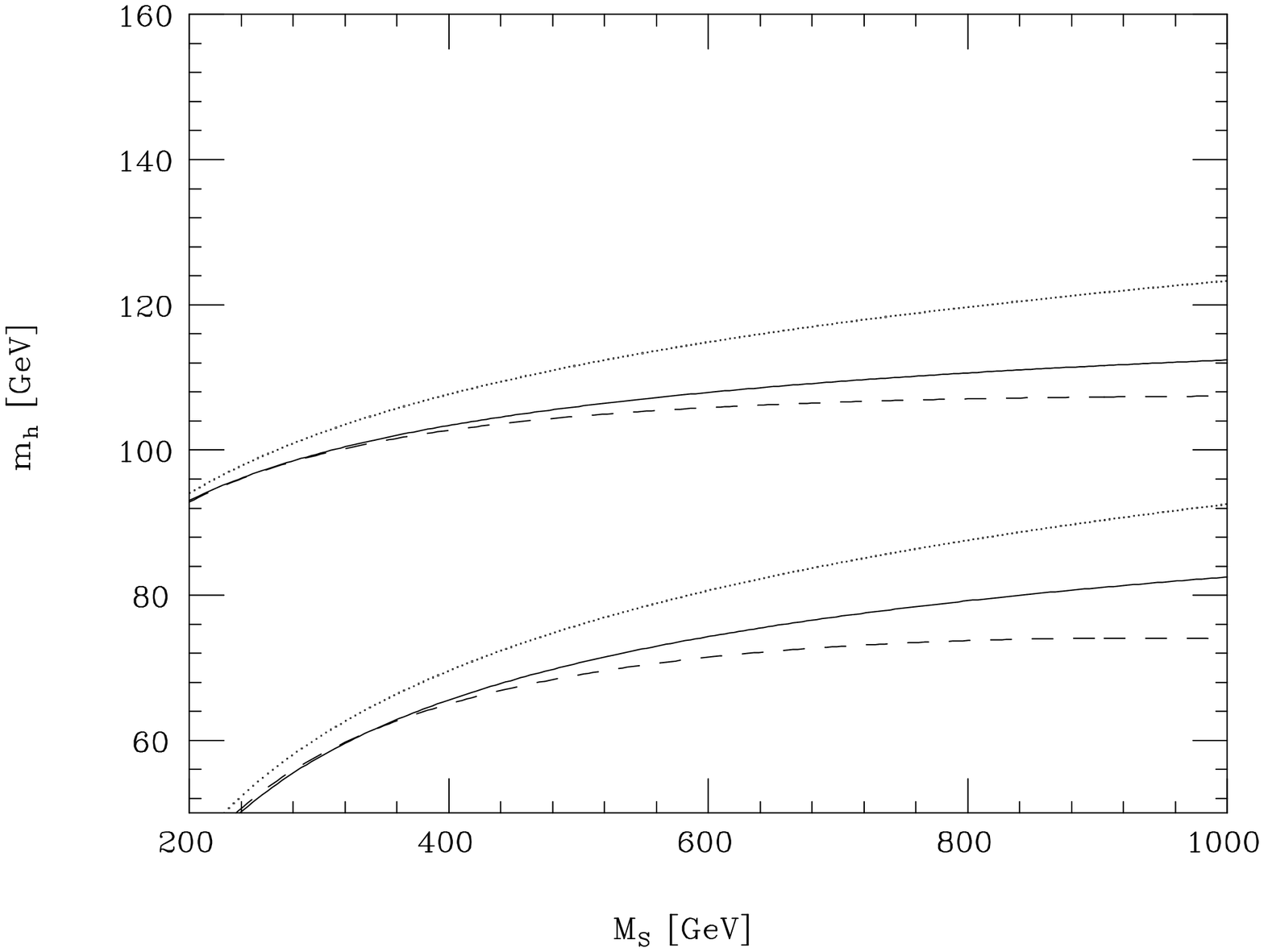,width=20cm,height=15cm,angle=90}}
\caption[0]{Plot of the Higgs mass $m_h$
from Eq.~(\ref{decmasa}) (solid lines), the
RG improved one-loop MSSM effective
potential (dashed lines) and the MSSM
effective potential considered at $Q^2=m_t^2$ (dotted lines),
as described in section 2,
for $M_t=175$ GeV,
$m_A = M_{S}$,
and $A_t = \mu =0$. The lower set
corresponds to
$\tan\beta=1.6$, and the upper set to $\tan\beta = 15$.}
\end{figure}
\begin{figure}
\centerline{
\psfig{figure=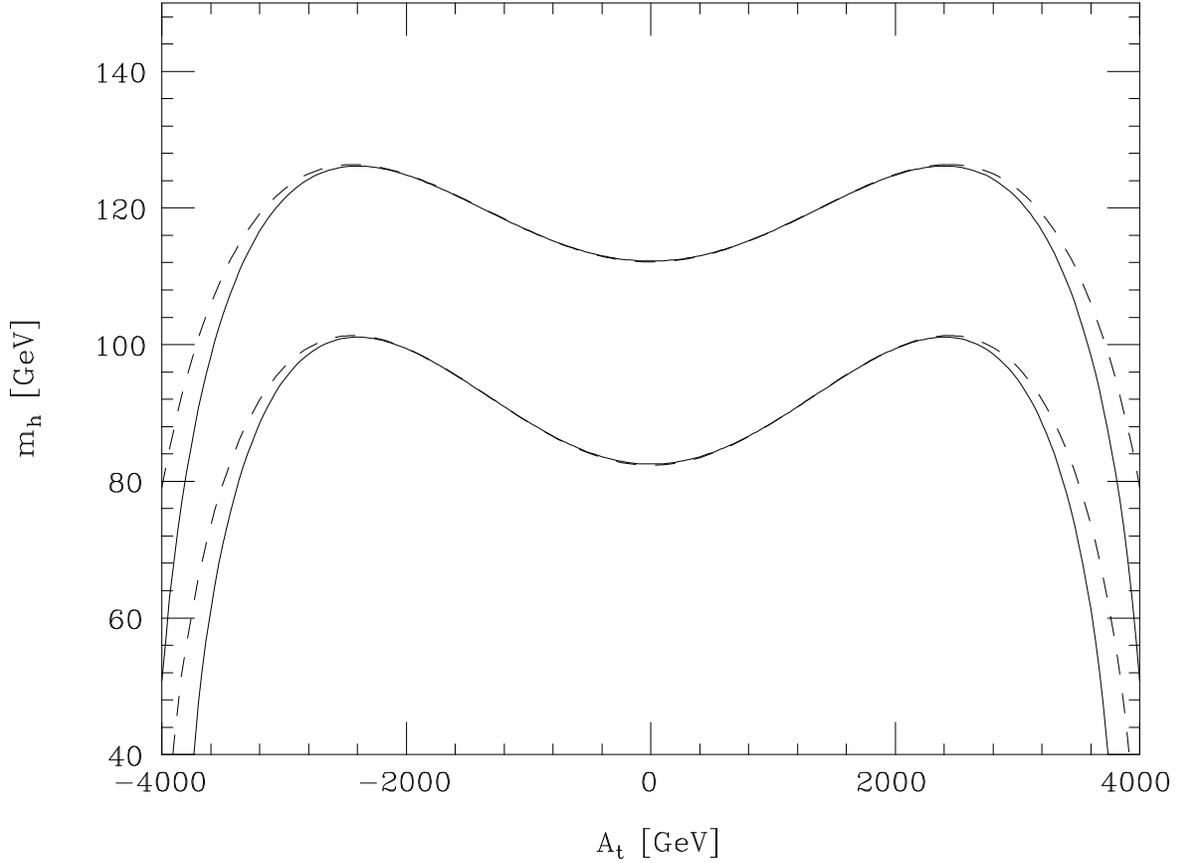,width=20cm,height=15cm,angle=90}}
\caption[0]{Plot of the Higgs mass $m_h$
as a function of $A_t$, from Eq.~(\ref{decmasa}),
using the exact threshold function (\ref{umbral})
(solid line) and the approximation (\ref{expumbral}) (dashed line),
for $M_t=175$ GeV, $\mu=0$
and $m_A=M_S=1$ TeV. The upper set corresponds to
$\tan\beta=15$ and the lower set to $\tan\beta=1.6$. }
\end{figure}
%
\begin{figure}
\centerline{
\psfig{figure=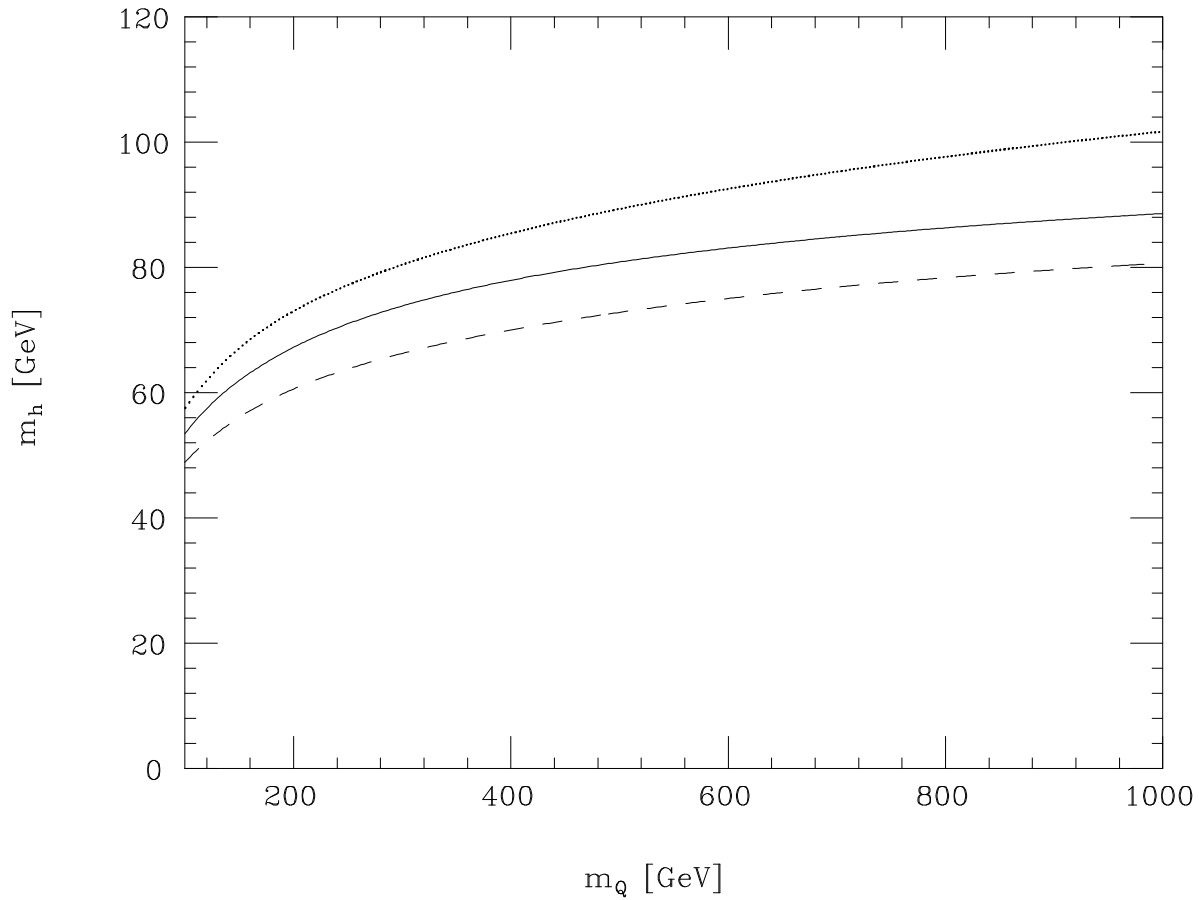,width=20cm,height=15cm,angle=90}}
\caption[0]{Plot of the Higgs mass $m_h$
as a function of $m_Q$ (solid line),
for $M_t=175$ GeV, $\tan\beta = 1.6$,
$\mu=A_b=0$, $m_U = m_D = 1$ TeV, and
values of the CP-odd Higgs mass $m_A$ and the stop
mixing mass parameter $A_t$ equal to $m_Q$. The
dashed and dotted lines denote the same as in Fig. 1. }
\end{figure}
%
\begin{figure}
\centerline{
\psfig{figure=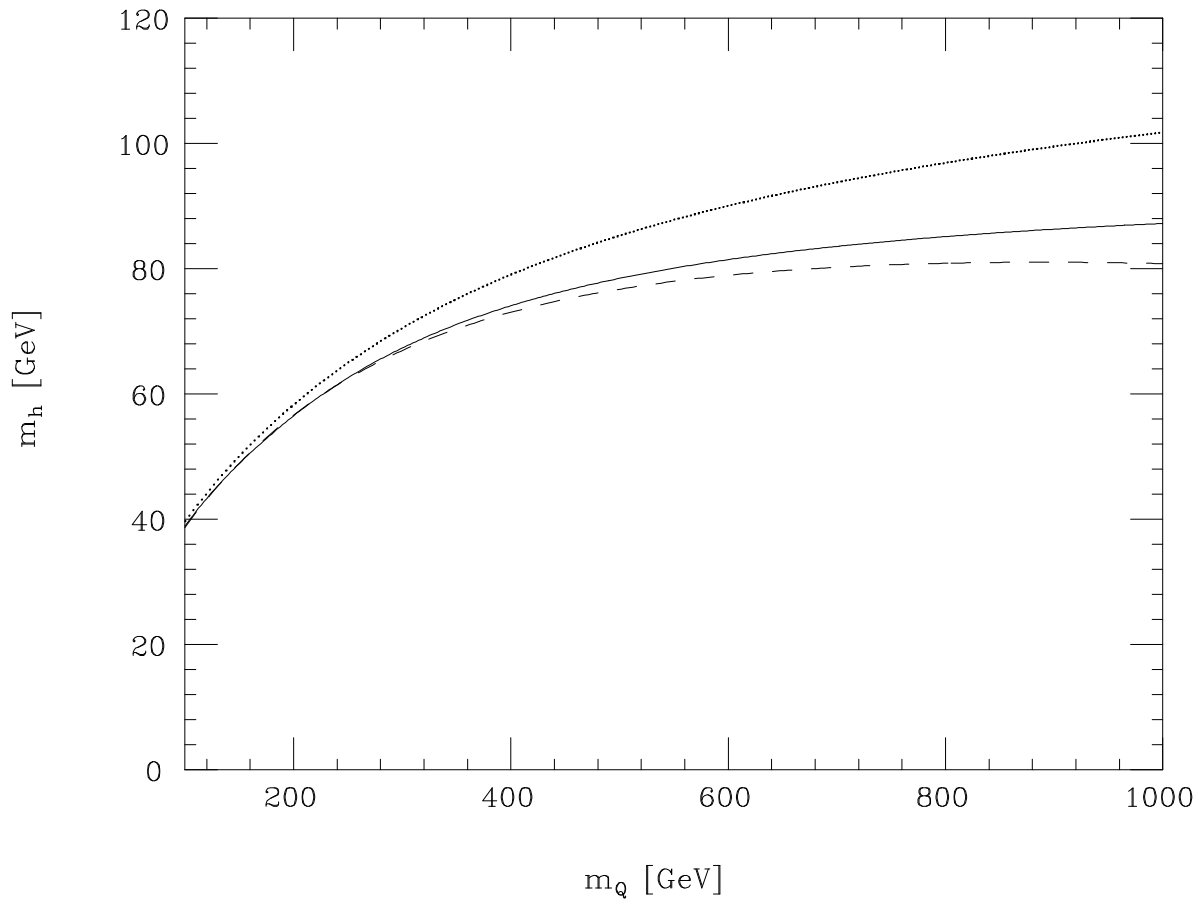,width=20cm,height=15cm,angle=90}}
\caption[0]{ The same as in Fig. 3, but for $m_U = m_D = 100$
GeV.}
\end{figure}
%
\begin{figure}
\centerline{
\psfig{figure=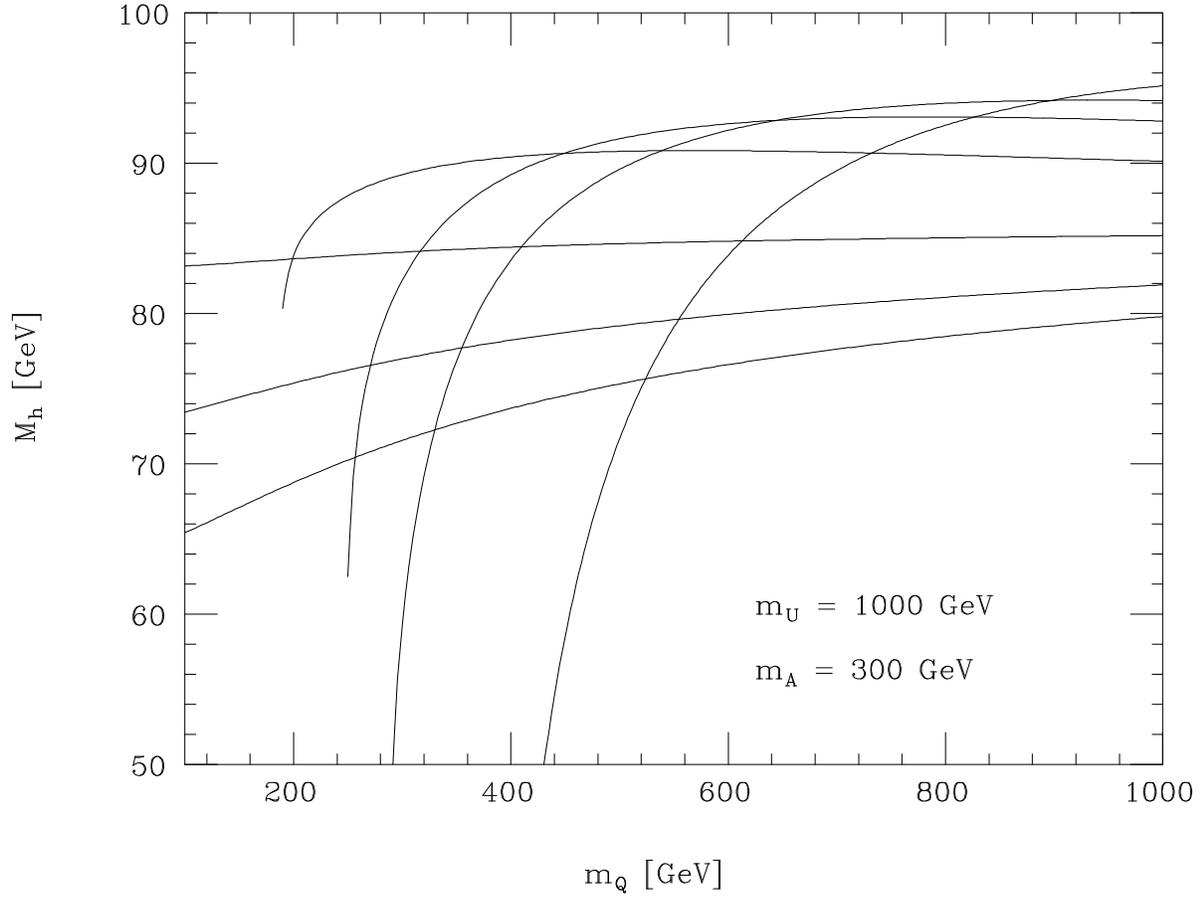,width=20cm,height=15cm,angle=90}}
\caption[0]{Plot of the pole Higgs mass $M_h$
as a function of $m_Q$, for $M_t=175$ GeV,
$\tan\beta=1.6$, $\mu=A_b=0$,
$m_U = m_D = 1000$ GeV
and $m_A = 300$ GeV. The different lines denote different
values of the $A_t$ parameter. Starting from below at
$m_Q = 1$ TeV, $A_t = 0$, 0.6, 1.0, 1.5,
1.8, 2.0  and
2.4 TeV, respectively.}
\end{figure}
%
\begin{figure}
\centerline{
\psfig{figure=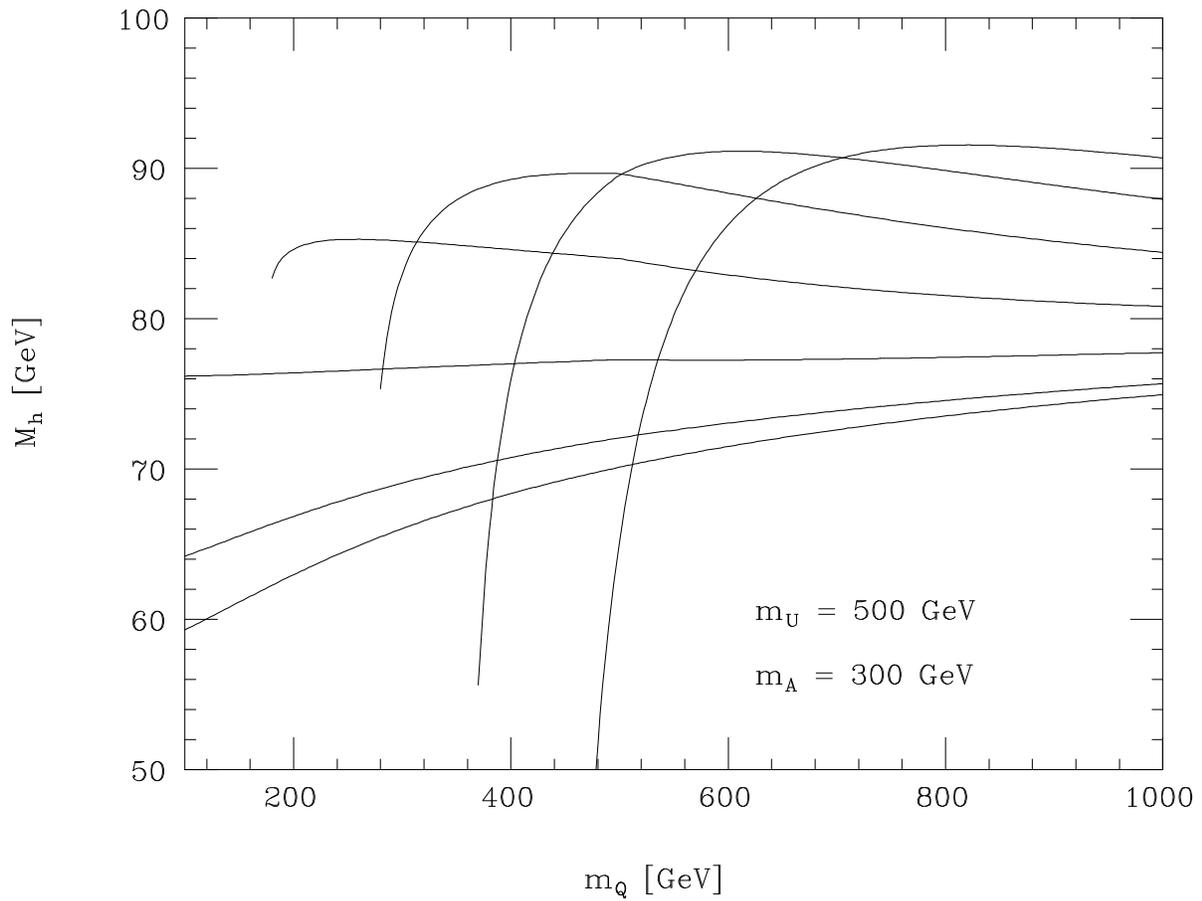,width=20cm,height=15cm,angle=90}}
\caption[0]{The same as Fig. 5 but for $m_U = 500$ GeV and
$A_t = 0$, 0.25, 0.5, 0.75, 1.0, 1.25  and 1.5 TeV, respectively.}
\end{figure}
%
\begin{figure}
\centerline{
\psfig{figure=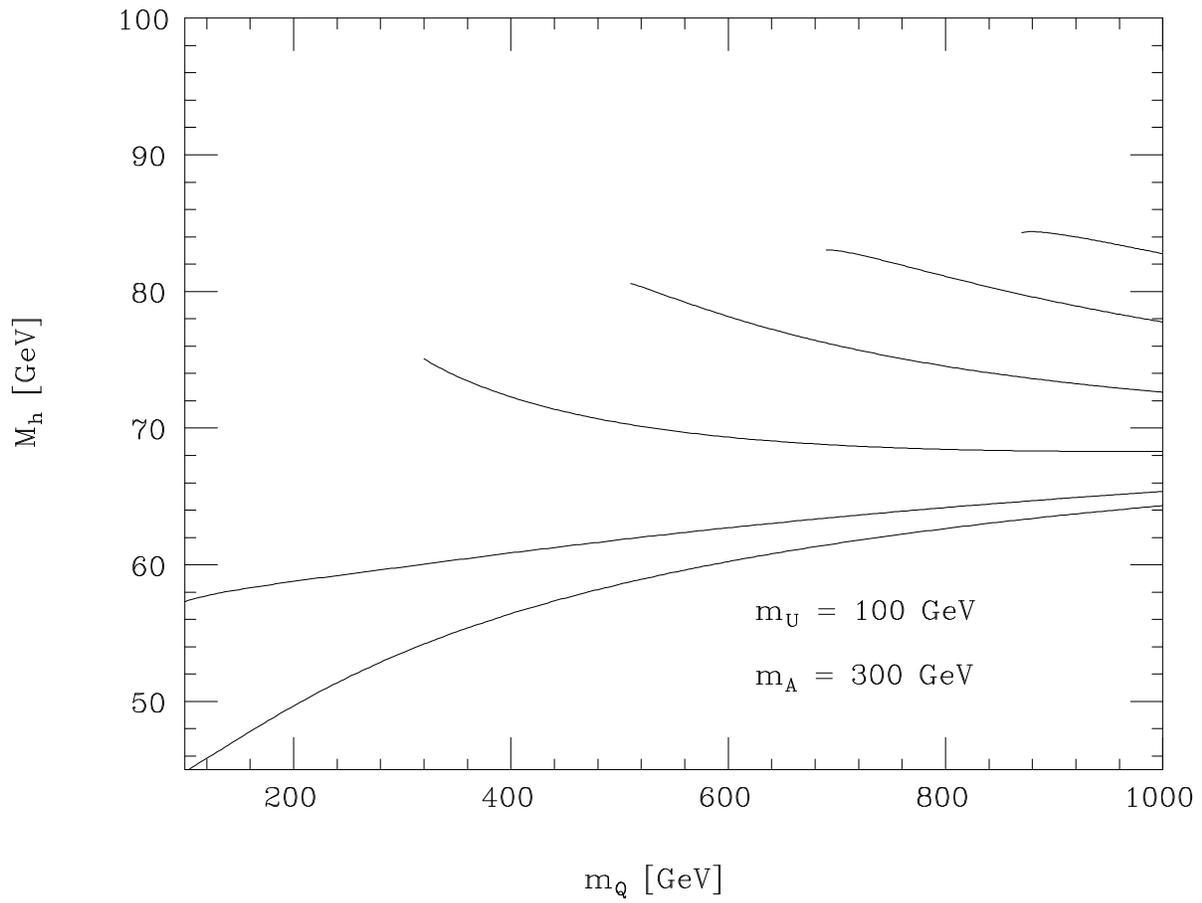,width=20cm,height=15cm,angle=90}}
\caption[0]{The same as Fig. 5 but for $m_U = 100$ GeV and
$A_t = 0$, 0.2, 0.4, 0.6, 0.8,  and 1 TeV, respectively.}
\end{figure}
%
\begin{figure}
\centerline{
\psfig{figure=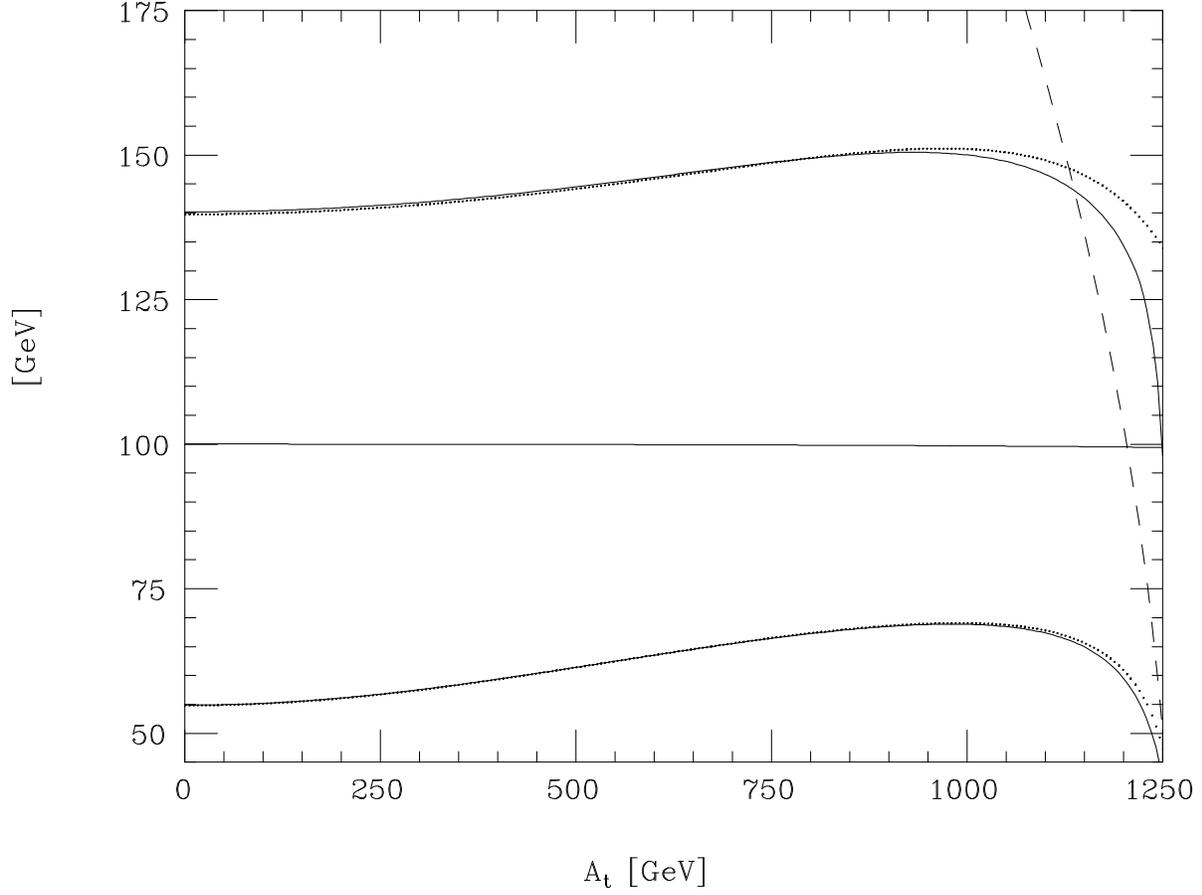,width=20cm,height=15cm,angle=90}}
\caption[0]{Plot of the pole (solid lines) and running (dotted
lines) Higgs masses for a running CP-odd Higgs mass
$m_A=100$ GeV, $\tan\beta=1.6$,
$m_Q=365$ GeV, $m_U=m_D=500$ GeV, $A_b=\mu=0$ and
$M_t=175$ GeV. The lower (upper) set corresponds to
the scalar Higgs boson $h$ ($H$), and middle curve to
the pseudoscalar Higgs boson $A$. The mass of the
lightest stop is plotted as the dashed
line. }
\end{figure}
%

\begin{figure}
\centerline{
\psfig{figure=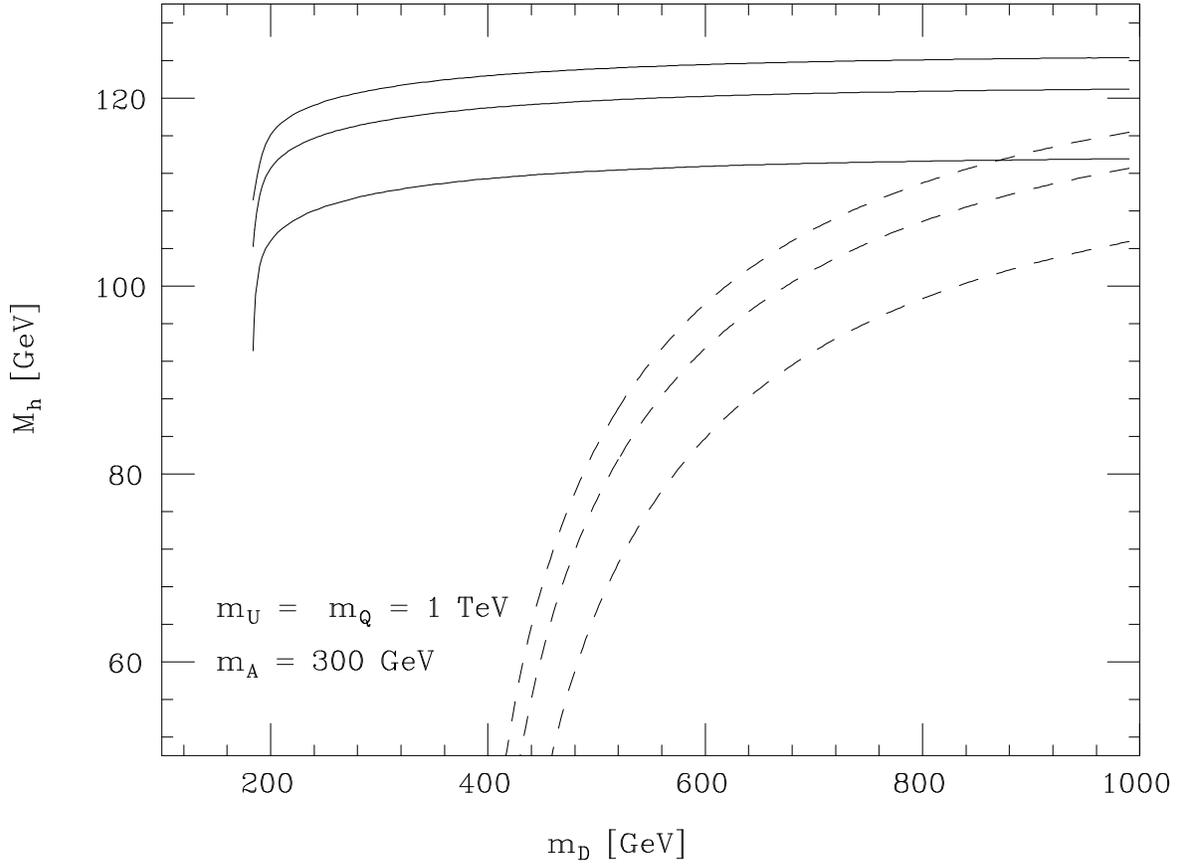,width=20cm,height=15cm,angle=90}}
\caption[0]{Plot of the pole  Higgs mass $M_h$
as a function of $m_D$, for $M_t=175$ GeV, $\tan\beta = 60$,
$A_b=0$, $m_Q = m_U = 1$ TeV,
$A_t=0,\; 1.5,\; 2.4$ TeV (from bottom to top) and
$\mu=1 $ TeV (solid curves) and $\mu=2$ TeV (dashed curves). }
\end{figure}
%

\end{document}